\documentclass[superscriptaddress,amsmath,amssymb,twocolumn,pra,floatfix]{revtex4-1}
\usepackage{amssymb,amsmath}
\usepackage{graphicx}
\usepackage{enumitem}
\usepackage[dvipsnames,usenames]{xcolor}
\usepackage[breaklinks=true,colorlinks=true,linkcolor=blue,urlcolor=blue,citecolor=blue]{hyperref}
\begin{document}

\title{Thermopower in a boundary driven bosonic ladder in the presence of a gauge field}

\author{Bo Xing}
\email{bo\_xing@mymail.sutd.edu.sg}
\affiliation{Science, Mathematics and Technology Cluster, Singapore University of Technology and Design, 8 Somapah Road, 487372 Singapore}
\author{Xiansong Xu}
\affiliation{Science, Mathematics and Technology Cluster, Singapore University of Technology and Design, 8 Somapah Road, 487372 Singapore}
\author{Vinitha Balachandran}
\affiliation{Science, Mathematics and Technology Cluster, Singapore University of Technology and Design, 8 Somapah Road, 487372 Singapore}
\author{Dario Poletti}
\affiliation{Science, Mathematics and Technology Cluster, Singapore University of Technology and Design, 8 Somapah Road, 487372 Singapore}
\affiliation{Engineering Product Development Pillar, Singapore University of Technology and Design, 8 Somapah Road, 487372 Singapore}

\date{\today}

\begin{abstract}
We consider a bosonic two-legged ladder whose two-band energy spectrum can be tuned in the presence of a uniform gauge field, to four distinct scenarios: degenerate or non-degenerate ground states with gapped or gapless energy bands.
We couple the ladder to two baths at different temperatures and chemical potentials and analyze the efficiency and power generated in the linear as well as nonlinear response regime.
Our results, obtained with the Green's function method, show that the maximum performance efficiency and generated power are strongly dependent on the type of the underlying energy spectrum.
We also show that the ideal scenario for efficient energy conversion, as well as power generation, corresponds to the case in which the spectrum has a gap between the bands, and the ground state is degenerate. 
\end{abstract}

\maketitle

\section{Introduction}\label{sec:intro}
Efficient energy harvesting is an important challenge faced by future technologies.
Thermoelectric conversion of work from heat offers a promising solution~\cite{Mahan1997, Dresselhaus2007, BenentiWhitney2017}.
However, thermodynamics places fundamental bounds on maximum efficiency and the generated power~\cite{Curzon1975EfficiencyOA, Benenti2011, Brandner2013, Whitney_2014}.
In linear response, the primary measure of the efficiency of thermoelectric devices or materials is the figure of merit $ZT=G S^2 T /K$, a function of temperature $T$, thermal conductance $K$, particle conductance $G$, and Seebeck coefficient $S$~\cite{Nolas2001ThermoelectricsBP, Heeaak9997, Snyder2017, Goldsmid2021}. 

Studies on energy harvesting have been focusing mainly on fermionic systems~\cite{Nolas2001ThermoelectricsBP, GoldsmidBook}, where the particles considered are typically electrons, although fermionic atoms in ultracold gases have been considered too~\cite{Brantut713}.

Studies on the thermopower performance of bosonic systems are still in their infancy compared to fermionic systems~\cite{Filippone2016, Papoular_2016, Gallego_Marcos_2014, Bidasyuk_2018, Oliveira2018}.
Bosonic systems, due to their uniquely defined Bose-Einstein distribution, can populate energy bands differently and may lead to novel insights in improving the thermopower performance.
Recent advances in cold atom experiments have greatly increased the ability to study transport for bosonic particles or excitations for example in ultracold gases experiments or with Josephson junctions~\cite{ChienDiVentra2015, FazioVDZant2001}.  
Experimental observation of transport phenomena of ultracold bosons has also been realised in 1D and quasi-1D systems~\cite{TanziModugno2013,SimpsonKruger2014,EckelHill2016,KrinnerEsslinger2017}.
In this paper, we aim to push forward the investigation of thermopower performance of bosonic systems.
In particular, we focus on a system that, even without any interactions, undergoes a quantum phase transition between a Meissner and a vortex phase~\cite{Kardar1986, Nishiyama2000, AtalaBloch2014}.
This allows us also to study the effect of quantum phase transitions on the thermopower performance of a bosonic system.    
For a review on transport in dissipatively boundary-driven systems and, in particular, the role of phase transitions, see~\cite{LandiSchaller2021}.

\begin{figure}[htp]
    \centering
    \includegraphics[width=\columnwidth]{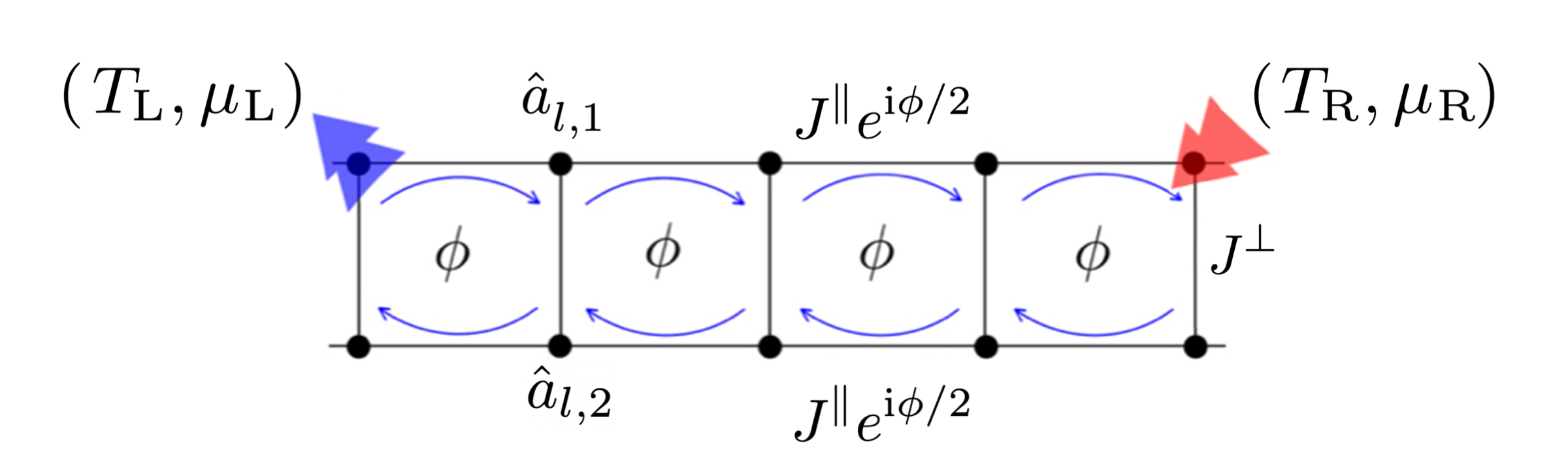}
    \caption{A schematic representation of our setup.
    The ladder consists of two coupled legs, with the bosonic creation (annihilation) operators described by $\hat{a}_{l,p}^{\dagger}$ ($\hat{a}^{}_{l,p}$), where $l=1,2\dots L$ refers to the site along the leg and $p=1,2$ refers to the top or bottom leg.
    $J^{\perp}$ and $J^{{\|}}$ are the tunneling amplitude along the rungs and legs of the ladder respectively.
    The magnetic field imposes a phase factor $\phi$ when hopping along the legs.
    The coupling to the left and right baths are represented by the blue and red double-arrow.
    The baths are coupled to the top corners of the ladder and are characterized by temperatures $T_{\mathrm{L}}$ or $T_{\mathrm{R}}$ and chemical potentials $\mu_{\mathrm{L}}$ or $\mu_{\mathrm{R}}$.
    For all results, we work in units for which $J^{\|}=k_B=\hbar=1$.} 
    \label{fig:model}
\end{figure}

The system we consider is a bosonic ladder with a uniform gauge field as shown in Fig.\ref{fig:model}.
A change in the gauge field may result in the ground state going from unique to degenerate, thus leading to a quantum phase transition between the Meissner and the vortex phases respectively.
It can also cause the opening of a gap in the two-band energy spectrum.
Hence, there are four qualitatively different energy spectrum structures in which the system can be tuned to. 
The influence of the energy spectrum on the transport properties of similar bosonic systems have been studied in~\cite{GuoPoletti2016, RivasMartin-Delgado2017, GuoPoletti2017, XingPoletti2020}. 

Here we investigate how tuning the system parameters to tailor the energy bands can be used to significantly alter its thermopower conversion performance~\cite{Mahan7436, Pei2012, Witkoske2017, Kumarasinghe2019, Rudderham2020, Zhou2011, Jeong2012}.
More in detail, we investigate the interplay between the boundary driving baths and the system parameters in tuning the heat-to-work conversion.
Using the non-equilibrium Green's function technique~\cite{CaroliSaint-James1971, MeirWingreen1992, HaugJauho2008, ProciukDunietz2010, Aeberhard2011, ZimbovskayaPederson2011, NikolicThygesen2012, DharHanggi2012, WangThingna2014, Ryndyk2016}, we focus on both linear and nonlinear response regimes.
To quantitatively evaluate the performance of the ladder, we use the figure of merit in the linear response regime and the efficiency and power generated in the nonlinear response regime.
In particular, we explore the four distinct regions in the system parameter space, each with a different type of energy band structure, and highlight the regions with the highest figure of merit, efficiency, or power generated. 

The paper is organized as follows: in Sec.~\ref{sec:model} we introduce the system and non-equilibrium setup, briefly describes the non-equilibrium Green's function, and introduce the Onsager coefficients used to study the system in the linear response regime.
We present the analysis on the figure of merit in the linear response regime in Sec.~\ref{sec:LRRresult} and study the efficiency and power generated away from the linear response regime in Sec. \ref{sec:NLRRresults}.
Lastly, we summarize our work in Sec.~\ref{sec:conclusions}.

\section{Model and Methods}\label{sec:model}
\subsection{Two-legged bosonic ladder}
We study a two-legged non-interacting bosonic ladder with a uniform gauge field.
The Hamiltonian of the ladder is
\begin{equation}\label{eqn:Hs}
    \begin{aligned}
        \hat{H}_{\rm S} = &-\left( J^{\|} \sum_{l,p} e^{i\left( -1 \right)^{p+1}\phi/2} \; \hat{a}_{l,p}^{\dagger} \hat{a}^{}_{l+1,p} \right. \\
                          &\left. + J^{\perp}\sum_{l} \hat{a}_{l,1}^{\dagger}\hat{a}^{}_{l,2} + \text{H.c} \right) + V\sum_{l,p}\hat{a}_{l,p}^{\dagger}\hat{a}^{}_{l,p}
    ,\end{aligned}
\end{equation}
where $\hat{a}_{l,p}^{\dagger}$ ($\hat{a}^{}_{l,p}$) is the bosonic creation (annihilation) operator at the $l$-th rung and $p$-th leg of the ladder, $J^{\|}$ ($J^{\perp}$) is the tunneling amplitude along the legs (rungs) and $V$ is the local potential.
Due to the presence of a gauge field, the bosons in the ladder acquire a phase $\phi$ when tunneling along the legs of the ladder.
The sign of the phase depends on the direction of the field circulation and is shown in Fig.~\ref{fig:model}.
In this article, we mainly consider a ladder with a length $L=64$ (128 sites)~\footnote{Simulations at $L=128$ have shown that the results obtained are consistent with $L=64$.} and a local potential $V/J^{\parallel}=8$.

\begin{figure}[htp]
    \centering
    \includegraphics[width=\columnwidth]{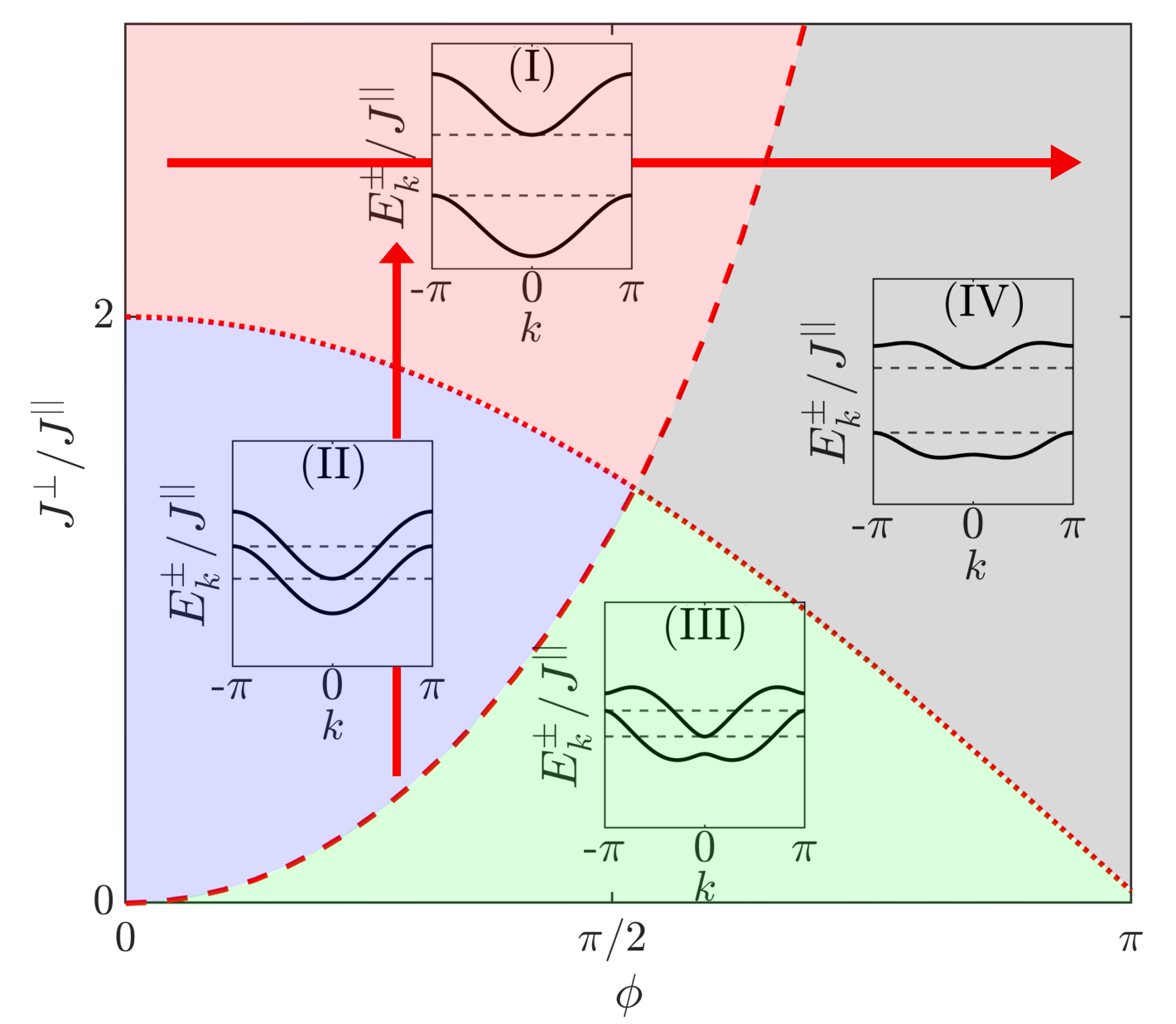}
    \caption{Energy band structures of a two-legged bosonic ladder in the system parameter space, $J^{\perp}/J^{\|}$ and $\phi$. The red dotted line, from Eq.~(\ref{eqn:line1}), and the dashed line, from Eq.~(\ref{eqn:line2}), divide the parameter space into four distinct regions I to IV.
    The Meissner (regions I and II) to vortex (regions III and IV) quantum phase transition takes place across the red dashed line.
    In each region, the band structure is noticeably different.
    For each region, we show an inset with the energy band structure, i.e., $E^{\pm}_{k}/J^{\|}$ versus the quasi momentum $k$, where $+$ and $-$ correspond to the upper and lower band respectively. In each inset, the dashed lines correspond to the energy levels of $\max( E^{-}_{k}/J^{\|} )$ and $\min( E^{+}_{k}/J^{\|} )$.
    In addition, two red arrows represent the locations of the horizontal and vertical cuts in the system parameter space which are studied in Figs.~\ref{fig:GSKZT_H},~\ref{fig:GSKZT_V},~\ref{fig:ZTbath}.}
    \label{fig:espec}
\end{figure}
The single-particle Hamiltonian in Eq.~(\ref{eqn:Hs}) with periodic boundary condition can be diagonalized readily and it has a two-band structure with energies $E^\pm_k$ with $k$ being the quasi momentum~\cite{Kardar1986}.
Depending on the magnitude of $J^{\perp}/J^{\|}$ and $\phi$, the energy spectrum of the ladder can be classified in four typical regions~\cite{GuoPoletti2016,XingPoletti2020}, as shown in Fig.~\ref{fig:espec}.
The red dotted line in Fig.~\ref{fig:espec}
\begin{equation}\label{eqn:line1}
    J_{c1}^{\perp} = 2J^{\|} \cos{\left( \phi/2 \right)}
,\end{equation}
gives the critical values of $J_{c1}^{\perp}$ at which the opening of the energy gap occurs.
The red dashed lines in Fig.~\ref{fig:espec}
\begin{equation}\label{eqn:line2}
    J_{c2}^{\perp} = 2J^{\|}\tan{\left( \phi/ 2\right)}\sin{\left( \phi/2 \right)}
\end{equation}
gives the critical values of $J_{c2}^{\perp}$ at which the degeneracy of the ground state occurs.
For $J^{\perp}>J_{c2}^{\perp}$ (regions I and II), the ground state of the ladder is in the Meissner phase, where the particle current only flows along the edges of the ladder.
For $J^{\perp}<J_{c2}^{\perp}$ (regions III and IV), the ground state of the ladder enters a vortex phase with finite inner rung currents.
The focus of our paper is to study how these quantum phases and their underlying energy band structure affect the performance of the system as an engine in both the linear and nonlinear response regimes.
In the following, we work in units for which $J^{\|}=k_B=\hbar=1$.

\subsection{Non-equilibrium setup}\label{sec:NEsetup}
We couple the ladder to two bosonic baths at different temperatures and chemical potentials at the top edges as shown in Fig.~\ref{fig:model}.
The baths are modeled as a collection of non-interacting bosons with Hamiltonian,
\begin{equation}
    \hat{H}_{\rm L/R} = \sum_{k} E^{}_{k,{\rm L/R}} \hat{b}_{k,{\rm L/R}}^{\dagger} \hat{b}^{}_{k,{\rm L/R}}.
\end{equation}
where $\hat{b}^{\dagger}_{k,{\rm L/R}}$ ($\hat{b}^{}_{k,{\rm L/R}}$) is the creation (annihilation) operator for a bosonic excitation with energy $E_{k,{\rm L/R}}$ in the left ($\mathrm{L}$) or right ($\mathrm{R}$) bath.

The baths are coupled to the system via the system-bath coupling Hamiltonian
\begin{align}
  \hat{H}_{I,{\rm L/R}} =& \sum_{k} c^{}_{k,{\rm L/R}} \left( \hat{a}_{{\rm L/R}}^{\dagger}\hat{b}^{}_{k,{\rm L/R}} + \hat{b}_{k,{\rm L/R}}^{\dagger}\hat{a}^{}_{{\rm L/R}} \right) \label{eqn:sbint}
,\end{align}
where $c_{k,\mathrm{L}/\mathrm{R}}$ denotes the coupling strength and $\hat{a}_{\rm L/R}^{\dagger},\;\hat{a}^{}_{\rm L/R}$ are the bosonic operators at the top edges of the ladder in contact with the baths.
Note that this choice of system-bath coupling conserves the total number of bosons for the overall system-plus-baths setup.

The baths are assumed to be at thermal equilibrium characterized by the Bose-Einstein distribution
\begin{equation}\label{eqn:boseEinstein}
    f(E,T,\mu) = \frac{1}{{e^{\left( E-\mu \right)/{T}} - 1}},
\end{equation}
at temperature $T = T_{\mathrm{L/R}}$ and chemical potential $\mu = \mu_{\mathrm{L/R}}$.
We fix the bath chemical potential $\mu_{\mathrm{L/R}}$ such that the ground state occupation of the bath is
\begin{align}\label{eqn:GSn0}
    \bar{n}_{0}\left( T_{\mathrm{L/R}},\mu_{\mathrm{L/R}} \right) &=\frac{1}{e^{\left( E_{0}-\mu_{\mathrm{L/R}} \right) / T_{\mathrm{L/R}}}-1} \\ \nonumber
                                                                  &=\frac{1}{e^{\tilde{E}_{\mathrm{L/R}}/ T_{\mathrm{L/R}}}-1},
\end{align}
where $\tilde{E}_{\mathrm{L/R}}$ is the ground state energy $E_{0}$ offset by the chemical potential $\mu_{\mathrm{L/R}}$, i.e. $\tilde{E}_{\mathrm{L/R}} = E_{0} - \mu_{\mathrm{L/R}}$.
In the following, we set the ground state occupation by fixing $\tilde{E}$.
In this way, we can better evaluate the role of energy band structure because the occupation of the excited states becomes dependent on the energy difference between the excited states and ground state at any given temperature.
In particular, we are interested in the scenario where the temperature bias competes with the chemical potential bias in driving a current.
This is achieved by choosing $T_{\mathrm{R}}>T_{\mathrm{L}}$ and $\mu_{\mathrm{L}}>\mu_{\mathrm{R}}$ (i.e. $\tilde{E}_{\mathrm{L}} < \tilde{E}_{\mathrm{R}}$).

\subsection{Green's function formalism}\label{sec:GreensF}
We use the non-equilibrium Green's function formalism~\cite{CaroliSaint-James1971, MeirWingreen1992, HaugJauho2008, ProciukDunietz2010, Aeberhard2011, ZimbovskayaPederson2011, NikolicThygesen2012, DharHanggi2012, WangThingna2014, Ryndyk2016} to study this non-equilibrium system-bath setup.
The retarded and advanced Green's function $G^{\rm r,a}(E)$ are
\begin{equation}
  G^{\rm r,a}(E)=\frac{1}{E-\hat{H}_{\rm S}-\Sigma_{\mathrm{L}}^{\rm r,a}(E)-\Sigma_{\mathrm{R}}^{\rm r,a}(E)}
,\end{equation}
where $\Sigma^{\rm r, a}_{\mathrm{L}/\mathrm{R}}\left( E \right)$ are the self-energy terms that model the effects of the baths on the isolated system.
$\Sigma^{\rm r, a}_{\mathrm{L}/\mathrm{R}}\left( E \right)$ is expressed in terms of the free Green's function of the baths $g^{\rm r,a}_{\mathrm{L}/\mathrm{R}} \left( E \right)=(E \pm i\epsilon- \hat{H}_{\mathrm{L}/\mathrm{R}})^{-1}$ and the coupling Hamiltonian $\hat{H}_{I,\mathrm{L}/\mathrm{R}}$,
\begin{equation}\label{eqn:selfenergy}
  \Sigma_{\mathrm{L}/\mathrm{R}}^{\rm r,a}\left( E \right) = \hat{H}^{}_{I,\mathrm{L}/\mathrm{R}}g_{\mathrm{L}/\mathrm{R}}^{\rm r,a}\left( E \right)\hat{H}_{I,\mathrm{L}/\mathrm{R}}^{\dagger}
.\end{equation}
The bath spectral density, or the level-width function,
\begin{align}
    \Gamma_{\mathrm{L}/\mathrm{R}}\left( E \right) & = {\rm i}(\Sigma^{\rm r}_{\mathrm{L}/\mathrm{R}} - \Sigma^{\rm a}_{\mathrm{L}/\mathrm{R}}) \nonumber \\
                                                   & = 2\pi \sum_{k} |c_{k,\mathrm{L}/\mathrm{R}}|^{2}\delta\left( E - E_{k,\mathrm{L}/\mathrm{R}} \right)
,\end{align}
characterizes the coupling between the system and baths.
We consider baths with Ohmic spectral density $\Gamma_{\mathrm{L}/\mathrm{R}}\left( E \right) = \gamma E $, where $\gamma$ is the effective system-bath coupling strength for each bath~\cite{DittrichZwerger1998}.

It follows that the particle current $\mathcal{J}_{P}$ and heat current $\mathcal{J}_{Q,L/R}$ are given by the Landauer-like formula~\cite{Landauer1957, Landauer1970}
\begin{equation}\label{eqn:pmeirwingreen}
    \mathcal{J}_{P} = \frac{1}{2\pi}\int_{-\infty}^{\infty} \!\!\! dE \; \mathcal{T}\left( E \right) \Lambda\left( E,T_{\mathrm{L,R}},\mu_{\mathrm{L,R}} \right)
,\end{equation}
\begin{equation}\label{eqn:hmeirwingreenL}
    \mathcal{J}_{Q,\mathrm{L}} = \frac{1}{2\pi}\int_{-\infty}^{\infty} \!\!\! dE \left( E - \mu_{\mathrm{L}} \right) \mathcal{T}\!\left( E \right) \Lambda\left( E,T_{\mathrm{L,R}},\mu_{\mathrm{L,R}} \right)
,\end{equation}
\begin{equation}\label{eqn:hmeirwingreenR}
    \mathcal{J}_{Q,\mathrm{R}} = -\frac{1}{2\pi}\int_{-\infty}^{\infty} \!\!\! dE \left( E - \mu_{\mathrm{R}} \right) \mathcal{T}\!\left( E \right) \Lambda\left( E,T_{\mathrm{L,R}},\mu_{\mathrm{L,R}} \right)
,\end{equation}
where $\mathcal{T}(E) = \mathrm{Tr}\left[ G^{\rm r}\left( E \right)\Gamma_{\mathrm{L}}\left( E \right)G^{\rm a}\left( E \right)\Gamma_{\mathrm{R}}\left( E \right)\right]$ is the transmission function~\cite{CaroliSaint-James1971} and $\Lambda\left( E,T_{\mathrm{L,R}},\mu_{\mathrm{L,R}} \right) = f{\left( E,T_{\mathrm{L}},\mu_{\mathrm{L}} \right)} - f{\left( E,T_{\mathrm{R}},\mu_{\mathrm{R}} \right)} $.
It is important to note that Eqs.~(\ref{eqn:pmeirwingreen},~\ref{eqn:hmeirwingreenL},~\ref{eqn:hmeirwingreenR}) are valid for two-terminal devices even when a magnetic field is present~\cite{Datta1995}.

While the particle currents entering and leaving the ladder are always the same in this non-equilibrium setup, the heat currents are only the same when the baths have the same chemical potential.
When the chemical potential is different, we can immediately observe from Eqs.~(\ref{eqn:hmeirwingreenL},~\ref{eqn:hmeirwingreenR}) that $\mathcal{J}_{Q,\mathrm{L}} \neq \mathcal{J}_{Q,\mathrm{R}}$.
For a multi-bath setup, the total power generated is the sum of all heat currents and it is given by 
\begin{equation}\label{eqn:power}
    \mathcal{P} = \sum_{i=\mathrm{L,R}} \mathcal{J}_{Q,i}.
\end{equation}
When $\mathcal{P}>0$, the system converts heat into work and act as an engine with an energy conversion efficiency quantified by
\begin{equation}\label{eqn:neng}
    \eta_{\mathrm{eng}}=\frac{\mathcal{P}}{\mathcal{J}_{Q,R}}
\end{equation} 
as shown, for instance, in \cite{BenentiWhitney2017}.
This expression is only valid when the currents $\mathcal{J}_{Q,R}$ and $\mathcal{J}_{Q,L}$ are respectively positive and negative which implies a heat flow from right to left, the scenario we study in this work.

\subsection{Thermopower in linear and non-linear response}\label{sec:LRR}
In the linear response regime, the currents are expanded to the linear order in biases $\Delta \mu = \left( \mu_{\mathrm{R}} - \mu_{\mathrm{L}} \right)$ and $\Delta T = \left( T_{\mathrm{R}}-T_{\mathrm{L}} \right)$ as~\cite{BenentiWhitney2017}
\begin{equation}\label{eqn:onsager}
    \begin{pmatrix}
        \mathcal{J}_{P} \\
        \mathcal{J}_{Q}
    \end{pmatrix}
    =
    \begin{pmatrix}
        \mathcal{L}_{PP} & \mathcal{L}_{PQ} \\
        \mathcal{L}_{QP} & \mathcal{L}_{QQ}
    \end{pmatrix}
    \begin{pmatrix}
        \Delta \mu  \\
        \Delta T /T
    \end{pmatrix}
,\end{equation}
where $T=\left( T_{\mathrm{L}}+ T_{\mathrm{R}}\right)/2$ is the average temperature. The elements of the $2\times 2$ matrix in Eq.~(\ref{eqn:onsager}) are the Onsager coefficients and can be fully determined in terms of the transmission coefficient $\mathcal{T}(E)$ as
\begin{equation}
    \mathcal{L}_{i,j}\!=\!\frac{1}{2\pi}\! \int_{-\infty}^{\infty} \!\!\! \frac{dE}{\hbar} \! 
    \begin{pmatrix}
        1 & E\! -\! \mu \\
        E\! -\! \mu & \left( E\! -\! \mu \right)^{2}
    \end{pmatrix}
        \mathcal{T}\!\left( E \right)\left[-f^{\prime}\!\left( E,T,\mu \right) \right]
,\end{equation}
where $f^{\prime}(E,T,\mu)$ is the derivative of the Bose-Einstein distribution, $\mu=\left( \mu_{\mathrm{R}} + \mu_{\mathrm{L}} \right)/2$ is the average chemical potential, and  $i,j=P,Q$.

The particle conductance, Seebeck coefficient, and thermal conductance are obtained from Eq.~(\ref{eqn:onsager}) as
\begin{eqnarray}
      G &=& \lim_{\Delta \mu \rightarrow 0}\frac{\mathcal{J}_{P}}{\Delta \mu}\Big |_{\Delta T = 0} = \mathcal{{L}_{PP}}, \label{eqn:G} \\
      S &=& -\lim_{\Delta T \rightarrow 0}\frac{\Delta \mu}{\Delta T}\Big |_{\mathcal{J}_{P} = 0} = \frac{1}{T}\frac{\mathcal{L}_{PQ}}{\mathcal{L}_{PP}}, \label{eqn:S}\\
      K &=&\lim_{\Delta T \rightarrow 0}\frac{\mathcal{J}_{Q}}{\Delta T}\Big |_{\mathcal{J}_{P} = 0} = \frac{1}{T}\left[\mathcal{L}_{QQ}-\frac{\mathcal{L}_{PQ}^2}{\mathcal{L}_{PP}}\right].\label{eqn:K}
    \end{eqnarray}

The thermopower performance of a material at a temperature $T$ is determined by the dimensionless figure of merit 
\begin{equation}\label{eqn:ZT}
   ZT = \frac{GS^{2}}{K}T
.\end{equation}
When the value of $ZT$ is higher, the energy conversion efficiency is higher. The maximum efficiency of a device can be quantified in terms of the single parameter $ZT$ as
\begin{equation}\label{eqn:etamax}
    \eta_{\max} = \eta_{\mathrm{C}}\frac{\sqrt{1+ZT}-1}{\sqrt{1+ZT}+1}
,\end{equation}
where $\eta_{\mathrm{C}} = 1 - T_{\mathrm{C}}/T_{\mathrm{H}}$ is the Carnot efficiency, $T_{\mathrm{C/H}}$ is the temperature of the cold and hot baths.
From Eq. (\ref{eqn:etamax}), it is clear that $ZT \to \infty$ leads to the Carnot efficiency.
Maximum power generated is another important quantity to characterize the thermopower performance and it is given by
\begin{equation}\label{eqn:Pmax}
    \mathcal{P}_{\max} = \frac{1}{4}S^{2}G\left( \Delta T \right)^{2}
.\end{equation}

When the difference in temperature and chemical potential of the two baths are finite, the particle and heat currents are highly nonlinear and expansion up to the linear order is not sufficient.
Hence, the analysis in the above Sec.~\ref{sec:LRR} does not apply.
However, it is possible to evaluate the power generated and the corresponding efficiency using Eqs.~(\ref{eqn:power},~\ref{eqn:neng}) numerically. 

\section{Results}\label{sec:Results}
In the following, we discuss the performance as thermopower converter of the two-legged ladder in the linear  (Sec.~\ref{sec:LRRresult}) and nonlinear response regimes (Sec.~\ref{sec:NLRRresults}).
Within the linear response, we analyze the engine efficiency of the four regions and explain the results in terms of the interdependencies of conductances and Seebeck coefficient.
In addition, we draw connections between the thermopower performance and the unique energy structure in each region.
We also investigate the role of system-bath coupling strength and chemical potential to improve efficiency.
Finally, we increase the biases and explore the nonlinear response of the two-legged ladder.

\subsection{Linear response regime}\label{sec:LRRresult}
\begin{figure}[htp]
    \centering
    \includegraphics[width=\columnwidth]{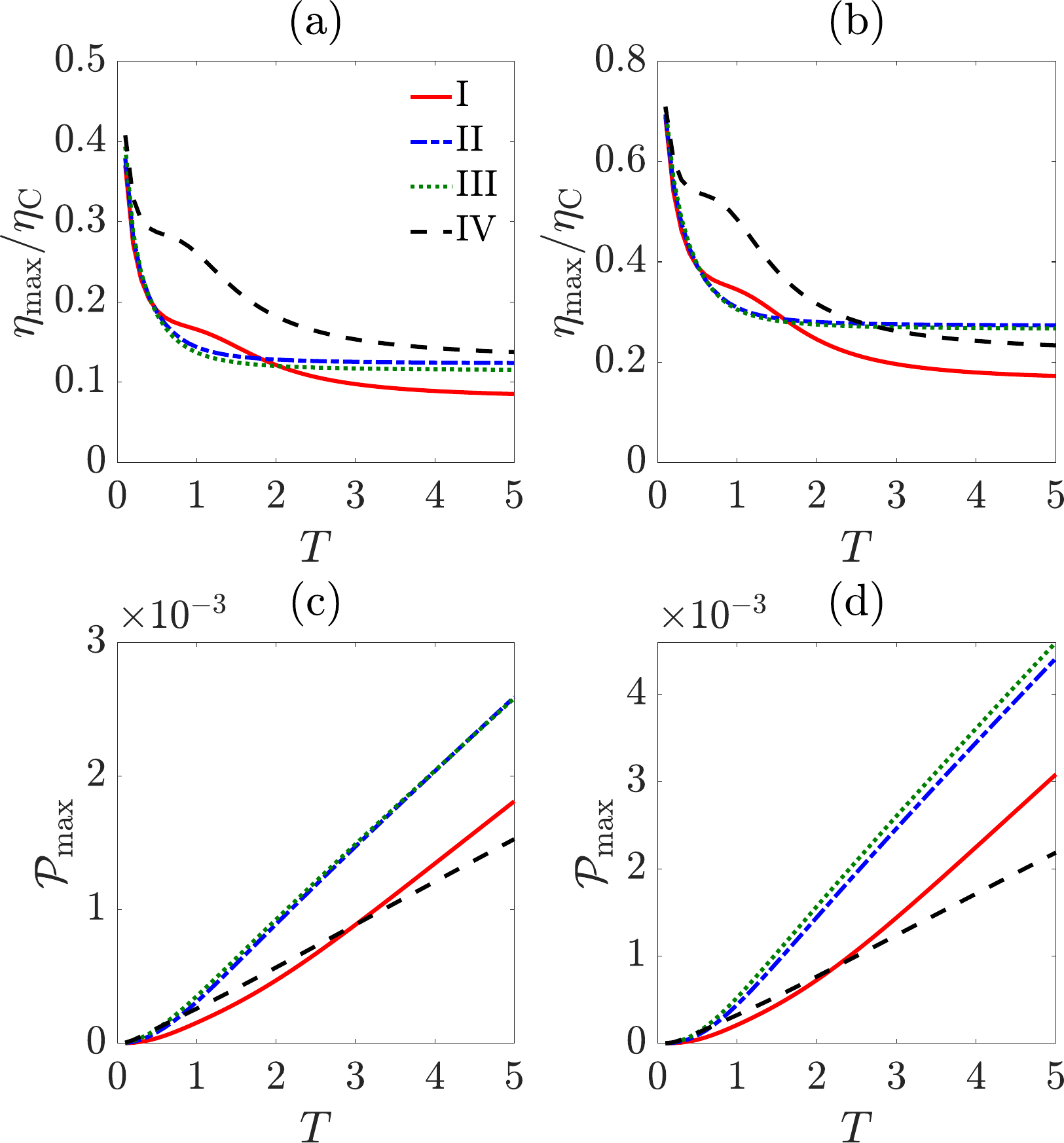}
    \caption{(a, b) Maximum efficiency (in terms of the Carnot efficiency), $\eta_{\max}/\eta_{\mathrm{C}}$, against average temperature $T$ for chemical potential $\mu = E_{0} - \tilde{E}$, where $\tilde{E} = 0.1$ and $0.5$ respectively.
    (c, d) Maximum power, $\mathcal{P}_{\max}$, against average temperature $T$ for $\tilde{E} = 0.1$ and $0.5$ respectively.
    A higher $\tilde{E}$ is equivalent to a lower $\mu$.
    To calculate $\mathcal{P}_{\max}$, we fix $\Delta T = 0.1T$.
    For each panel, the four lines represent different regions as described in the legend. 
    The $J^\perp$ and $\phi$ chosen for the red lines is $J^\perp=4.0$, $\phi=0.1\pi$ (region I).
    For the blue lines, $J^\perp=1.0$, $\phi=0.3\pi$ (region II).
    For the green lines, $J^\perp=1.0$, $\phi=0.5\pi$ (region III).
    For the black lines, $J^\perp=3.0$, $\phi=0.8\pi$ (region IV).
    For all panels, $\gamma=0.1$.}
    \label{fig:EFF_P_LRR}
\end{figure}
We start by studying the maximum efficiency (in terms of the Carnot efficiency), $\eta_{\max}/\eta_{\mathrm{C}}$, and maximum power, $\mathcal{P}_{\max}$, using the linear response theory for the four regions shown in Fig.~\ref{fig:espec}.
For each region, we choose arbitrary combinations of system parameters $\mathcal{J}^{\perp}$ and $\phi$ (one from each region) to represent the general behavior of the region.
We note that choosing another set of $\mathcal{J}^{\perp}$ and $\phi$ within the same region results in small quantitative changes in the observables we study.
However, the qualitative behavior of these regions does not show any significant dependence on the choice of the parameters.

In Fig.~\ref{fig:EFF_P_LRR}, we investigate the effect of average temperature, $T$, on $\eta_{\max}/\eta_{\mathrm{C}}$ and $\mathcal{P}_{\max}$ for two different average chemical potentials, $\mu = E_{0} - \tilde{E}$ where $\tilde{E} = ( \tilde{E}_{\mathrm{L}} + \tilde{E}_{\mathrm{R}} )/2$.
Two interesting observations stand out immediately when comparing Fig.~\ref{fig:EFF_P_LRR}(a, b).
Firstly, the behavior of regions I and IV, where the energy bands are gapped, are noticeably different from regions II and III at intermediate temperatures $0.5<T<1.0$.
More specifically, instead of decreasing rapidly to an asymptotic value, the maximum efficiency $\eta_{\max}/\eta_{\mathrm{C}}$ plateaus in this intermediate $T$ region, before decreasing further.
Secondly, comparing Fig.~\ref{fig:EFF_P_LRR}(a, b), we note that when chemical potential, $\mu$, is large (low $\tilde{E}$, panel (a)), the efficiency is smaller in all regimes.
For large chemical potential, region IV is always the most efficient regime within the temperature $T$ range we have explored.
At low chemical potential (high $\tilde{E}$, panel (b)), the most efficient region changes from region IV to region II as temperature increases.

In Fig.~\ref{fig:EFF_P_LRR}(c, d), we study the maximum power generated at different $T$ when $\Delta T = 0.1T$.
It is clear from the figure that power generated is negligible at low temperatures and increases monotonously as temperature rises.
Analyzing the panels we see that some regions generate more power than the rest depending on the temperature.
While region IV seems to deliver the most power at very low $T$, it is quickly overtaken by regions II and III as $T$ increases.
At high $T$, regions II and III, where the energy bands overlap, produce a substantially higher $\mathcal{P}_{\max}$ than regions I and IV.
Comparing Fig.~\ref{fig:EFF_P_LRR}(c, d), we find that the decrease in $\mu$ (increase in $\tilde{E}$) boosts the power generated and does not change the behavior of $\mathcal{P}_{\max}$ versus $T$.

\begin{figure}[htp]
    \centering
    \includegraphics[width=\columnwidth]{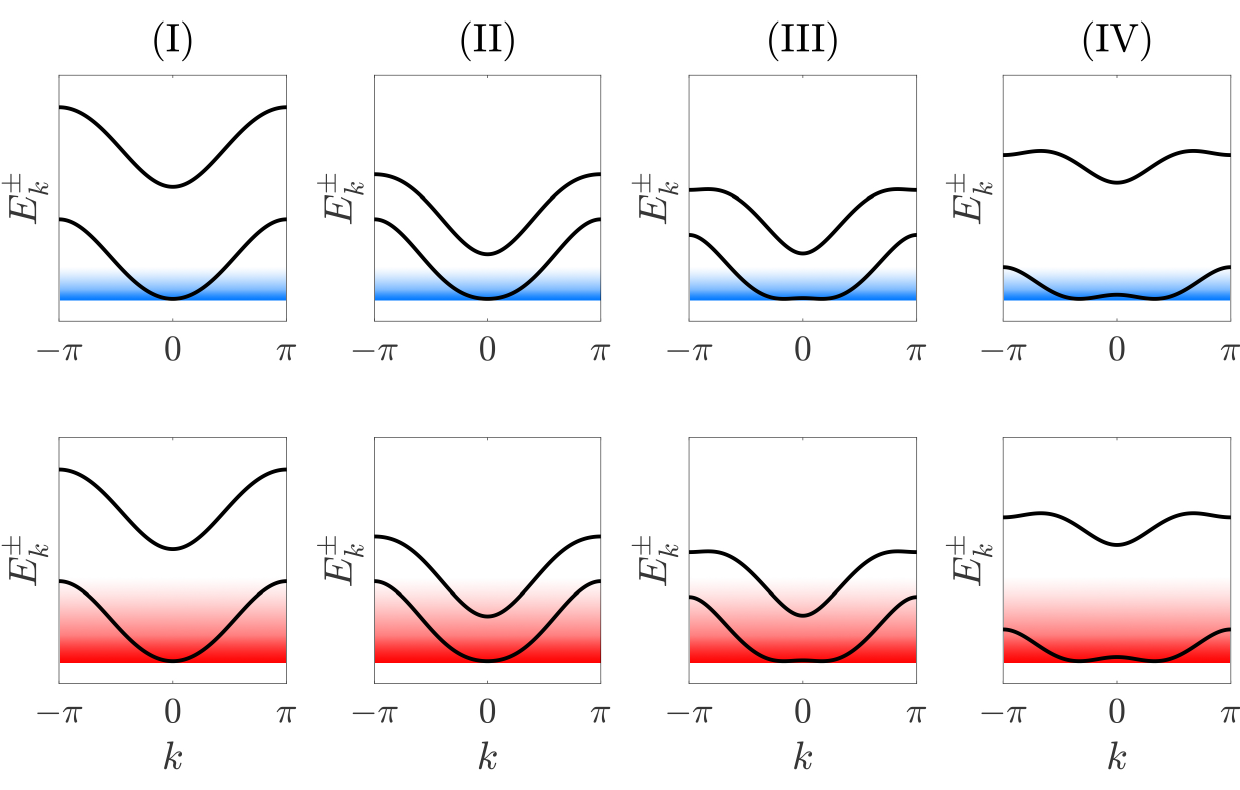}
    \caption{Band structure of the system in the four different regions. 
    The band structure of each region is characterized by the presence or absence of a bandgap and/or degenerate ground states.
    In each panel, the colored shading (blue and red) represents the occupation of the system at a given energy $E$.
    The blue (first row) represents a scenario where $T=0.1$ and the red (second row) represents $T=1.0$.
    At higher temperatures, the occupation of the states with higher energy becomes non-negligible.}
    \label{fig:bandstructure}
\end{figure}

To better understand the results of Fig.~\ref{fig:EFF_P_LRR}, in particular the difference in efficiency of each region, we evaluate the band structure of each region under different bath temperatures. 
In Fig.~\ref{fig:bandstructure}, we plot the band structure of each region. 
In each panel, the colored shading (blue and red) represents the occupation of the energy states.
The blue shading (first row) represents $T=0.1$ and the red shading (second row) represents $T=1.0$. 
The energy states in the system are filled very differently depending on the underlying band structure.
At a low temperature (first row), both particle and thermal transport are dominated by the low energy states.
This is true for all regions.
Therefore, the principal factor differentiating the regions is the density of energy states in the vicinity of the ground state.
Since regions III and IV have degenerate ground states, their lower bands are narrower and they have more energy states in the close vicinity of the ground state.
At this temperature, the presence of a bandgap does not influence the transport properties of the system, because the higher energy states are not occupied.

When the temperature is raised (second row), the particle transport is still dominated by the low energy states due to the nature of Bose-Einstein distribution, Eq.~(\ref{eqn:boseEinstein}).
However, thermal transport is influenced by the non-negligible presence of the higher energy states.
These higher energy states do not contribute to particle transport significantly, but play an important role in thermal transport due to the high energy they carry.
Therefore, we can expect the bandwidth of the lower band and the bandgap to influence thermal transport at higher temperature.
When the bands are not gapped and temperature is high, energy states from the upper band, or higher energy states from the lower band, can be substantially occupied and contribute to thermal transport.
However, as the bandgap opens, or when the bandwidth of the lower band becomes narrower, the higher energy states become inaccessible, resulting in a reduction in thermal transport.

\begin{figure}[htp]
    \centering
    \includegraphics[width=\columnwidth]{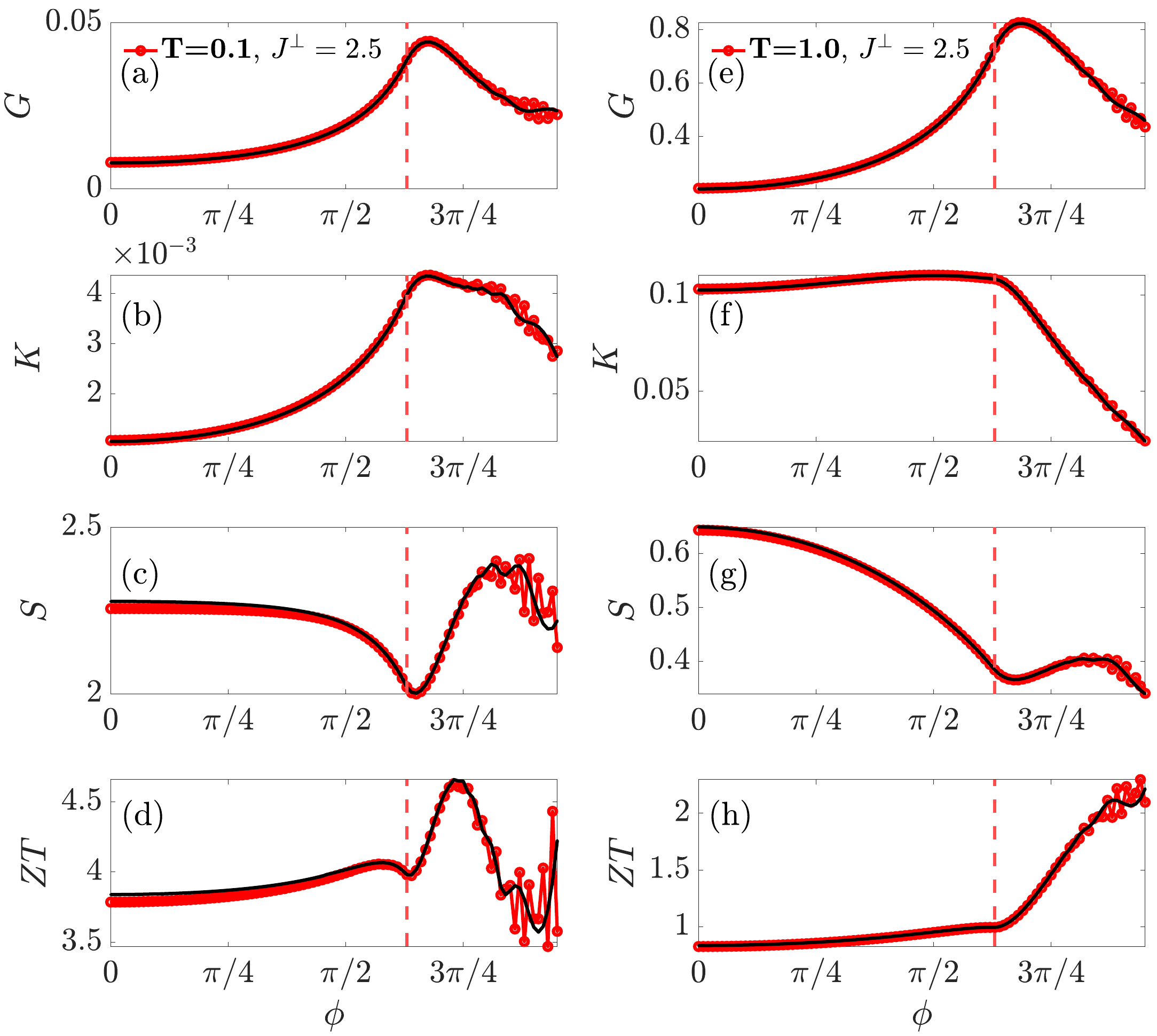}
    \caption{Particle conductance, $G$, thermal conductance, $K$, Seebeck coefficient, $S$, and figure of merit $ZT$ against $\phi$, when $J^{\perp} = 2.5$.
    This is a horizontal cut in the system parameter space and features a change from region I to IV as $\phi$ increases.
    A red horizontal arrow representing the location of this cut is shown in Fig.~\ref{fig:espec}.
    On the left column, $T=0.1$, and on the right column, $T=1.0$.
    The vertical dashed line signals the emergence of degenerate ground states (vortex phase).
    For all panels, $\gamma=0.1$ and $\mu = E_{0} - \tilde{E}$, where $\tilde{E} = 0.1$.
    The fluctuations in the plots are finite-size effects.
    As the system size increases, the amplitude and frequency of the fluctuation decrease.
    The black solid line is obtain from $L=128$.}
    \label{fig:GSKZT_H}
\end{figure}

To exemplify the analysis above, we demonstrate the particle conductance, $G$, thermal conductance, $K$, Seebeck coefficient, $S$, and figure of merit $ZT$ as a function of $\phi$.
In Fig.~\ref{fig:GSKZT_H}, we perform a horizontal cut across the system parameter space at $J^{\perp}=2.5$ to evaluate the effects that come with the emergence of degenerate ground states.
This horizontal cut features a change from region I to IV as $\phi$ increases and can be visualised in Fig.~\ref{fig:espec}.
The dashed vertical line marks the location where the transition takes place.
Region I is on the left of the line and region IV is on the right.

We find that the emergence of degenerate ground states greatly changes the transport properties at both low and high temperatures.
At $T=0.1$ (left column), $G$ and $K$ peak right after the emergence of degenerate ground states while $S$ reaches a minimum.
We relate this to the emergence of degenerate ground states which greatly increases the number of low energy particles participating in the transport.
As $\phi$ continues to increase beyond the transition, the degenerate ground states become further apart in the momentum space and number of energy states close to the ground stat starts to decreases after reaching a peak.
At $T=1.0$ (right column), the behavior of $G$ is similar to $T=0.1$.
This is not surprising as $G$ is dominated by the lower energy states, which make up the majority of the particles regardless of the temperature.
However, the behavior of $K$ and $S$ are very different.
$K$ and $S$ are heavily influenced by the higher energy states, which have non-negligible occupations only at higher temperature.
The narrowing of the lower band after the transition results in a reduction in the availability of higher energy states.
As a result, we find that $K$ decreases even more rapidly after the transition at higher temperature.
This leads to a significant increase in $ZT$ after the transition.
It is worth pointing out that the fast fluctuation in the figures are finite-size effects. In fact the amplitude and frequency of the oscillation decrease when increasing the system size. 
This is shown in Fig.~\ref{fig:GSKZT_H} in which we plotted the data for $L=64$ with the red line with circles and for $L=128$ with the black continuous line.

\begin{figure}[htp]
    \centering
    \includegraphics[width=\columnwidth]{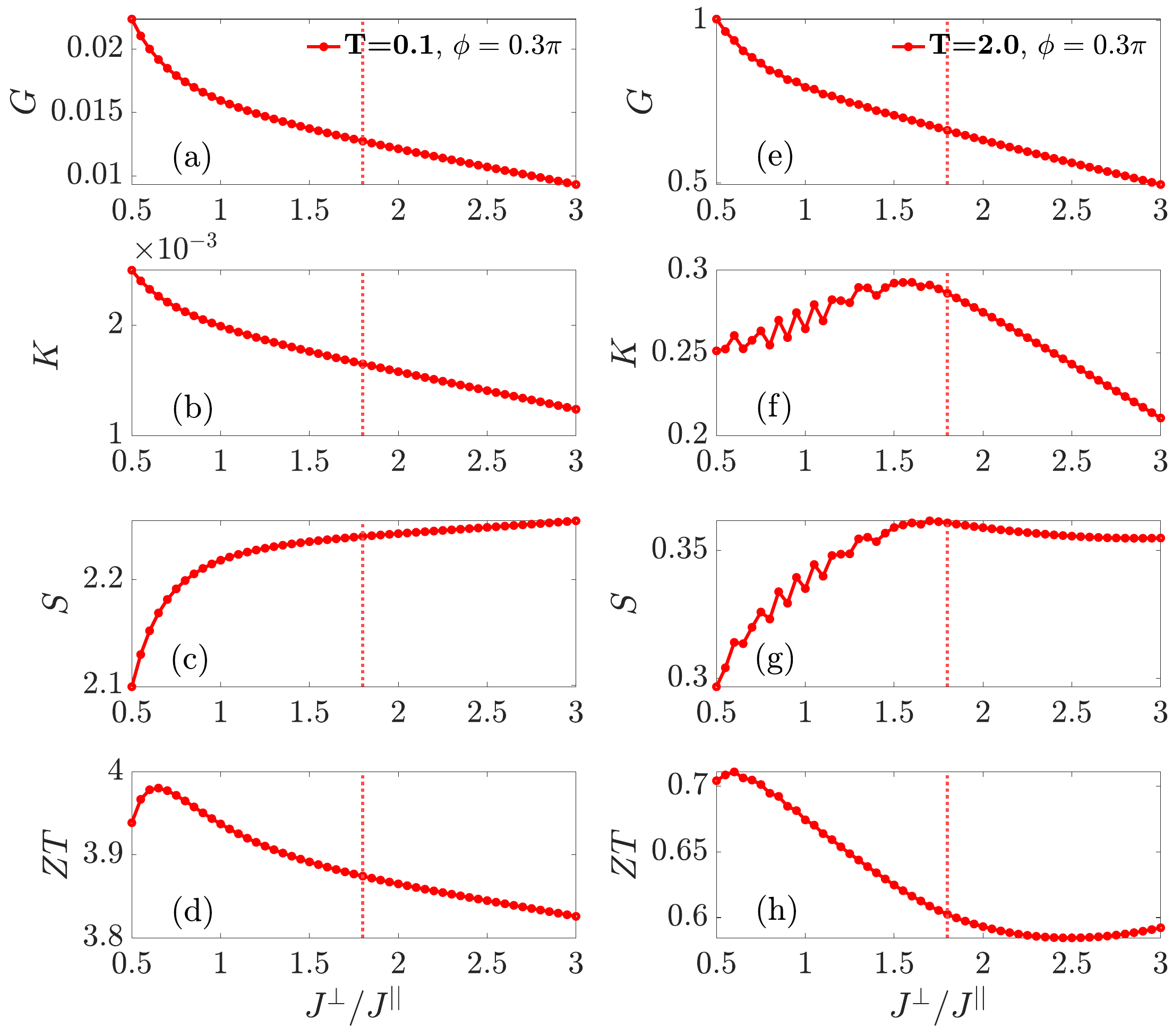}
    \caption{Particle conductance, $G$, thermal conductance, $K$, Seebeck coefficient, $S$, and figure of merit $ZT$ against $J ^{\perp}/J^{||}$, when $\phi = 0.3\pi$.
    This is a vertical cut in the system parameter space and features a change from region II to I as $J^{\perp}$ increases.
    A red vertical arrow representing the location of this cut is shown in Fig.~\ref{fig:espec}.
    On the left column, $T=0.1$, and on the right column, $T=2.0$.
    The vertical dotted line signals the opening of a bandgap between the upper and lower bands.
    For all panels, $\gamma=0.1$ and $\mu = E_{0} - \tilde{E}$, where $\tilde{E} = 0.1$.}
    \label{fig:GSKZT_V}
\end{figure}

In Fig.~\ref{fig:GSKZT_V}, we perform a vertical cut across the system parameter space at $\phi=0.3\pi$ to evaluate the effects that come with the opening of a bandgap.
We focus on the part of the cut which features a change from region II to I as $J^{\perp}$ increases.
The location of this vertical cut can be visualised in Fig.~\ref{fig:espec}.
The dotted vertical line marks the location where the gap between the two bands opens.
Region II is on the left of the line and region I is on the right.

In Fig.~\ref{fig:GSKZT_V}, we find that the opening of the bandgap only affects the thermal transport properties at high temperatures. 
At $T=0.1$ (left column), the opening of the bandgap does not impact the transport properties of the system.
As mentioned previously, particle and thermal transport are dominated by lower energy states at low temperatures.
The opening of the bandgap is irrelevant because the occupation of the high energy states are negligible.
When the temperature is substantially higher at $T=2.0$ (right column), we see that $G$ behaves similar to $T=0.1$.
However, the non-negligible occupation of higher energy states at $T=2.0$ gives rise to a totally different behavior for $K$.
In particular, $K$ peaks right before the opening of the band gap and falls rapidly after.

\begin{figure}[htp]
    \centering
    \includegraphics[width=\columnwidth]{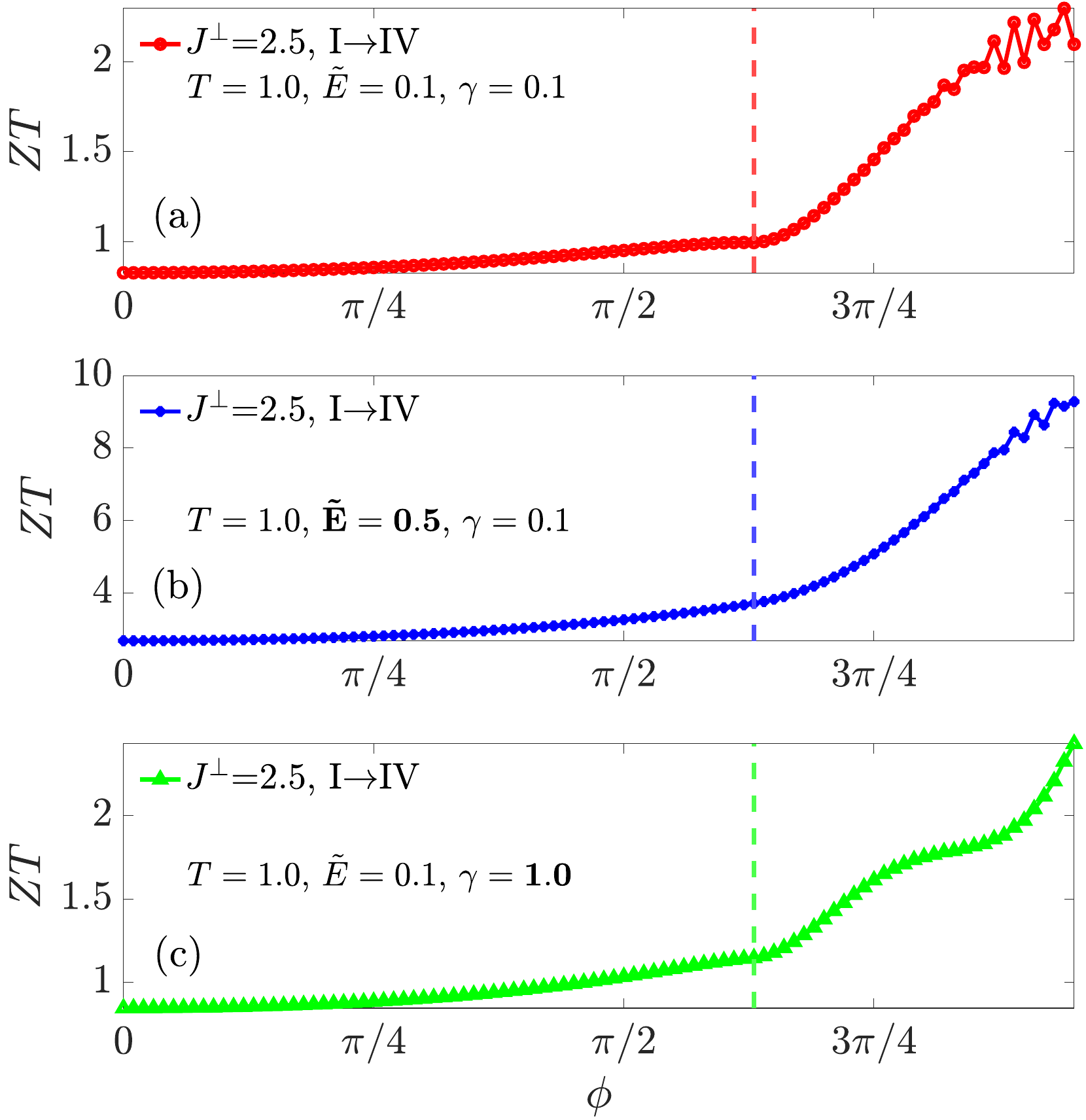}
    \caption{Figure of merit, $ZT$ against $\phi$ for different bath and system-bath setups.
    Similar to Fig.~\ref{fig:GSKZT_H}, the horizontal cut is at $J^{\perp} = 2.5$ and features a transition from region I to IV.
    The vertical dashed line signals the formation of degenerate ground states (Meissner to vortex phase).
    In the first panel, $T=1.0, \gamma=0.1$ and $\mu = E_{0} - \tilde{E}$, where $\tilde{E} = 0.1$.
    The subsequent panels feature a change in either $\gamma$ or $\tilde{E}$, where the change is highlighted in bold.}
    \label{fig:ZTbath}
\end{figure}

In the following, we focus on how the system-bath coupling, $\gamma$, and the choice of chemical potential, $\mu = E_{0} - \tilde{E}$, affect $ZT$ in different regions.
We study the setup with three different sets of bath and system-bath parameters in Fig.~\ref{fig:ZTbath}.
For all panels, we plot the horizontal cut of the system parameter space at $J^{\perp} = 2.5$, which features a transition from region I to region IV (Meissner to vortex transition), while the band gap is always open.
We stress that, despite the changes in the bath and system-bath coupling, the energy band structure continues to play an important role in determining transport performance.
Specifically, we see that $ZT$ increases significantly after the emergence of a degenerate ground states, regardless of the change in bath and system-bath parameters.
In (a), we study the $ZT$ of the setup at $T = 1.0, \tilde{E} = 0.1, \gamma = 0.1$ and find a monotonous increase in $ZT$ after the emergence of ground state degeneracy.
When $\tilde{E}$ increases from $0.1$ to $0.5$ in (b), we observe that $ZT$ increases significantly for all $\phi$ values.
The increase is especially remarkable as the system undergoes a transition from region I to IV.

\begin{figure*}[t]
    \centering
    \includegraphics[width=\linewidth]{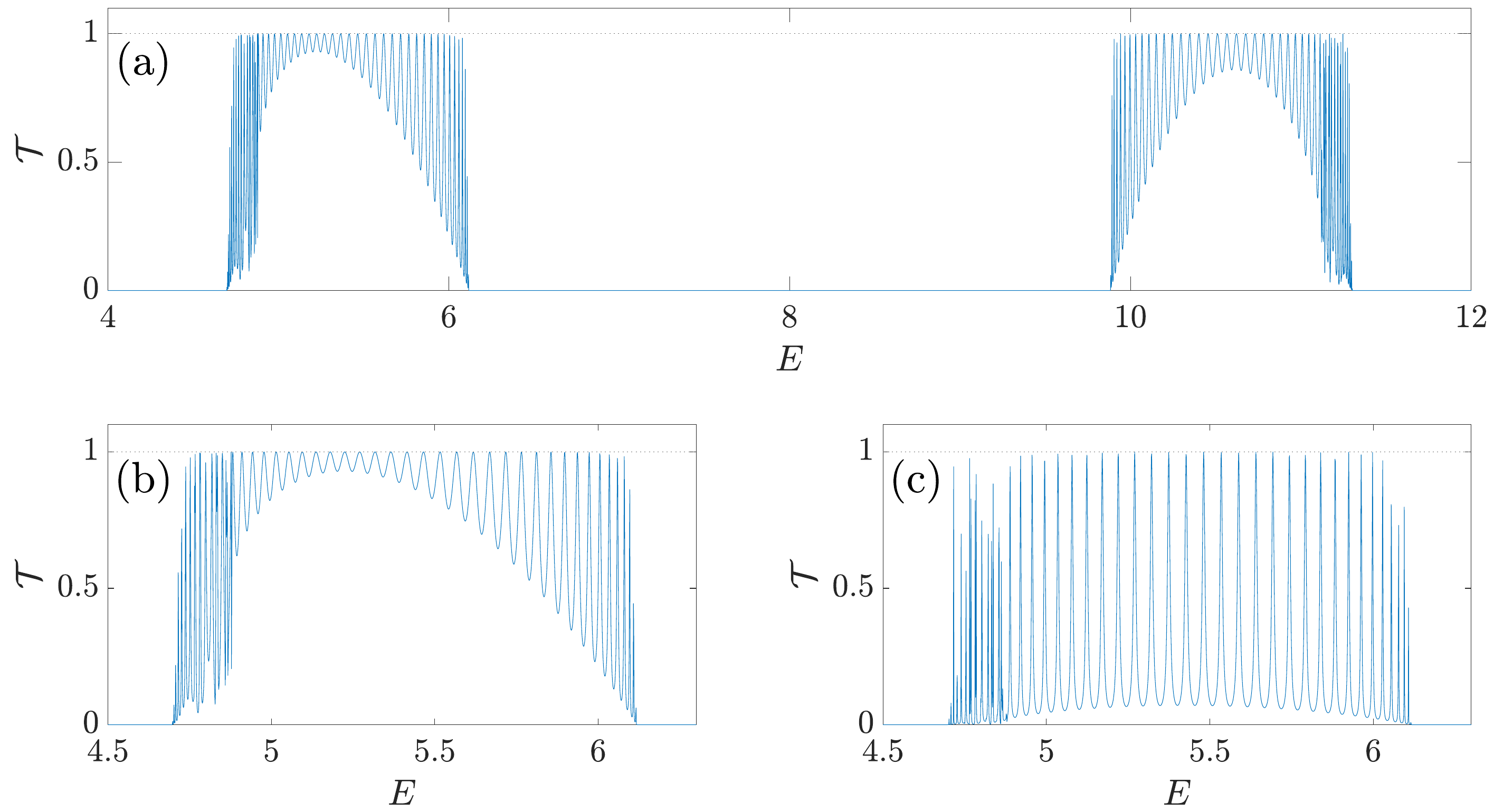}
    \caption{The transmission function, $\mathcal{T}(E)$, against energy $E$.
    The system parameter is $J^{\perp}=2.5, \phi=0.8\pi$ (region IV).
    In (a), we plot the transmission function for system-bath coupling $\gamma=0.1$.
    In (b), we plot the same transmission function for $\gamma = 0.1$, but zoom into the energy range of the lower band.
    In (c), we plot the transmission function of the lower band for $\gamma = 1.0$.
    A different combination of $J^{\perp}$ and $\phi$ in the same region does not present qualitatively different findings.
    In all panels, a $\mathcal{T}(E)=1$ dotted line is shown to demonstrate that the transmission function max out at $1$.} 
    \label{fig:transmission}
\end{figure*}

The effect of $\gamma$ on $ZT$ comes entirely from the transmission function, $\mathcal{T}(E)$, which has a prominent role in Eqs.~(\ref{eqn:pmeirwingreen}-\ref{eqn:hmeirwingreenR}). 
In Fig.~\ref{fig:transmission}, we examine $\mathcal{T}(E)$ at $J^{\perp} = 2.5$ and $\phi = 0.8\pi$ (region IV) for (a) $\gamma = 0.1$, (b) $\gamma = 0.1$ and (c) $\gamma = 1.0$~\footnote{We benchmarked our transmission function against the ones obtained from Kwant~\cite{groth_kwant_2014}, a widely used and cited python package for quantum transport, and found perfect agreement.}.
In (b) and (c), we show only the transmission function of the lower energy band.
The increase in $\gamma$ results in noticeable changes in the shape of the transmission function. 
In particular, the peaks becomes narrower and the minima of the transmission are lower.
However, as demonstrated in Fig.~\ref{fig:ZTbath}(c), the qualitative behavior of $ZT$ remains unchanged.

\subsection{Nonlinear response regime}\label{sec:NLRRresults}
Linear response theory gives indications on the performance of each region at some average temperature and chemical potential when the temperature and chemical potential biases are small.
When these biases are large, the evaluation of thermopower performance using linear response theory becomes invalid.
For such scenarios, we evaluate the efficiency and power generated directly using Eqs.~(\ref{eqn:power},~\ref{eqn:neng}). 

\begin{figure}[htp]
    \centering
    \includegraphics[width=\columnwidth]{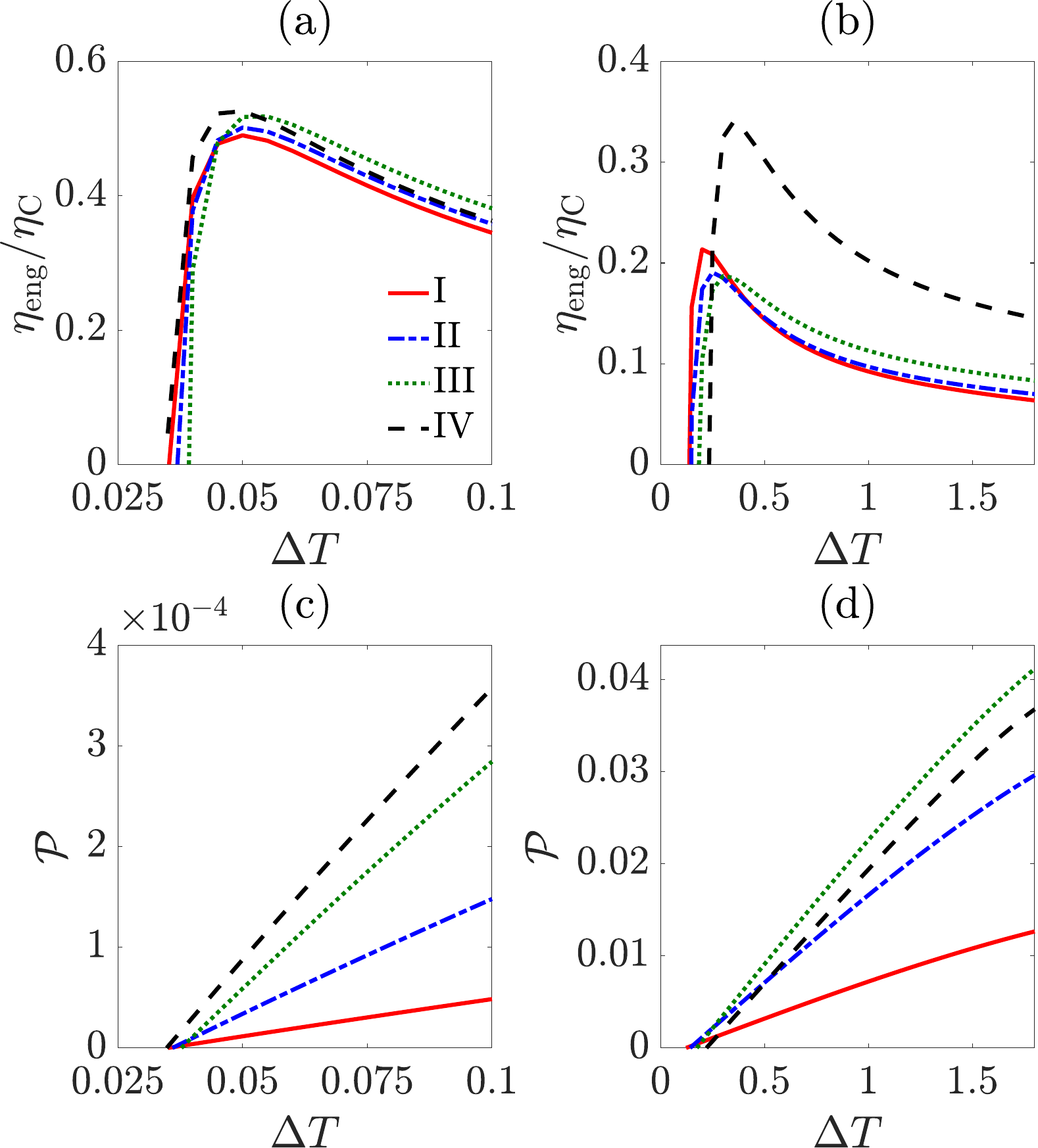}
    \caption{Efficiency (in terms of the Carnot efficiency) $\eta_{\mathrm{eng}}/\eta_{\mathrm{C}}$ (a,b) and power generated $\mathcal{P}$ (c,d) against temperature bias $\Delta T$.
    In (a,c), $T=0.1$ and in (b,d), $T=1.0$
    For each panel, the four lines represent different regions as described in the legend.
    The $J^\perp$ and $\phi$ chosen for the lines are identical to the description in Fig.~\ref{fig:EFF_P_LRR}.
    For all panels, $\gamma = 0.1$ and $\mu_{\mathrm{L/R}} = E_{0} - \tilde{E}_{\mathrm{L/R}}$, where $\tilde{E}_{\mathrm{L}}=0.1$ and $\tilde{E}_{\mathrm{R}} = 0.2$.}
    \label{fig:Eff_P}
\end{figure}

In Fig.~\ref{fig:Eff_P}, we plot the efficiency (in terms of the Carnot efficiency), $\eta_{\mathrm{eng}}/\eta_{\mathrm{C}}$, and power generated, $\mathcal{P}$, of the ladder as a function of $\Delta T = \left( T_{\mathrm{R}}-T_{\mathrm{L}} \right)$.
For Fig.~\ref{fig:Eff_P}(a, c), we fix the average temperature $T=\left( T_{\mathrm{L}}+T_{\mathrm{R}} \right)/2 = 0.1$, and for Fig.~\ref{fig:Eff_P}(b, d), $T=1.0$.
For all panels, $\mu_{\mathrm{L}} = E_{0} - \tilde{E}_{L}$ and $\mu_{\mathrm{R}} = E_{0} - \tilde{E}_{\mathrm{R}}$, where $\tilde{E}_{\mathrm{L}} = 0.1$ and $\tilde{E}_{\mathrm{R}} = 0.2$.
We find that the efficiency is maximum at some intermediate $\Delta T$ for all regions in Fig.~\ref{fig:Eff_P}(a, b).
All regions have similar $\eta_{\mathrm{eng}}/\eta_{\mathrm{C}}$ at $T=0.1$ (a), and region IV has a much higher maximum efficiency than other regions at $T=1.0$ (b).
This is qualitatively similar to our findings in the linear response regime, where we find region IV to be the most efficient at $T=1.0$ due to the presence of both the bandgap and degenerate ground states.
As $T$ increases from $T=0.1$ (a) to $1.0$ (b), the maximum efficiency of the ladder is reduced in all regions.
This reduction of maximum efficiency in all regions when $T=0.1 \to 1.0$ is predicted in the linear response regime as well, where we observe that $ZT$ is a function that decreases with $T$.

In Fig.~\ref{fig:Eff_P}(c, d), we plot the power generated by the four regions when $T= 0.1$ and $1.0$ respectively.
In general, the power generated $\mathcal{P}$ increases with both the increase in $T$ and $\Delta T$ as showed in panels Fig.~\ref{fig:Eff_P}(c, d).
In addition, the region that generates the most power depends on the $T$ it operates at.
The most efficient region does not necessarily generate the most power.
At $T=0.1$, region IV generates more power than all other regions.
However, when $T=1.0$, region III overtakes as it is gapless and hence can populate the higher energy states more efficiently, increasing the heat current $\mathcal{J}_{Q}$.

\begin{figure}[htp]
    \centering
    \includegraphics[width=\columnwidth]{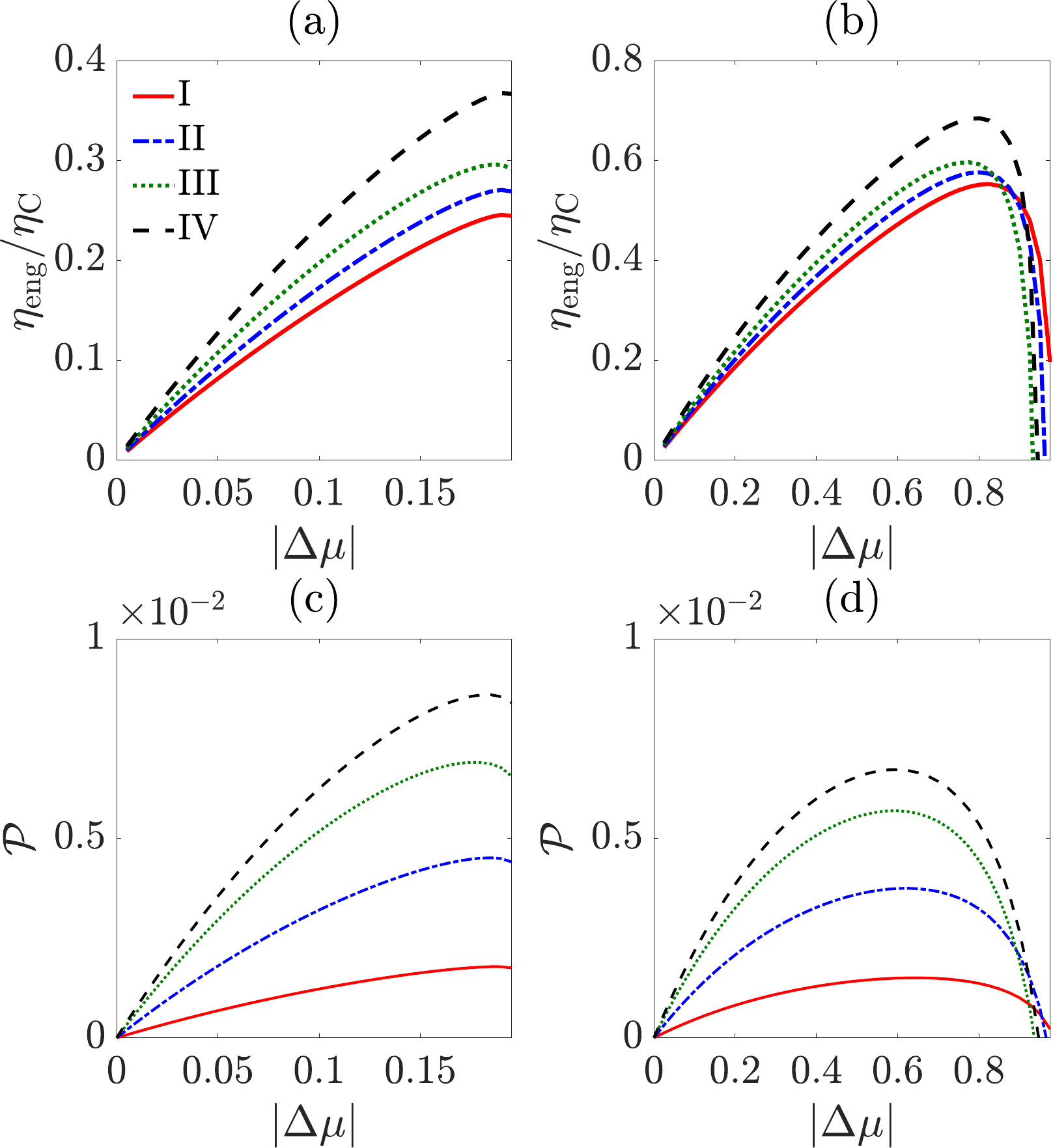}
    \caption{Efficiency (relative to Carnot's efficiency) $\eta_{\mathrm{eng}}/\eta_{\mathrm{C}}$ (a, b) and power generated $\mathcal{P}$ (c, d) against the strength of chemical potential bias, $|\Delta \mu|$.
    In (a, c), the average chemical potential $\mu = E_{0} - \tilde{E}$, where $\tilde{E} = 0.1$.
    In (b, d),  $\tilde{E}=0.5$.
    For each panel, the four lines represent different regions as described in the legend. 
    The $J^\perp$ and $\phi$ chosen for the lines are identical to the description in Fig.~\ref{fig:EFF_P_LRR}.
    For all panels, $T_{\mathrm{L}}=0.1$, $T_{\mathrm{R}}=0.5$ and $\gamma=0.1$.}
    \label{fig:Eff_P_mu}
\end{figure}
In Fig.~\ref{fig:Eff_P_mu}, we plot the efficiency (in terms of the Carnot efficiency), $\eta_{\mathrm{eng}}/\eta_{\mathrm{C}}$, and power generated, $\mathcal{P}$, of the ladder against the strength of chemical potential bias, $|\Delta \mu| = |\mu_{\mathrm{R}} - \mu_{\mathrm{L}}| $.
For Fig.~\ref{fig:Eff_P_mu}(a, c), we fix the average chemical potential $\mu = E_{0} - \tilde{E}$, where $\tilde{E} = 0.1$.
For Fig.~\ref{fig:Eff_P_mu}(b, d), $\tilde{E}=0.5$.
For all panels, $T_{\mathrm{L}}=0.1$ and $T_{\mathrm{R}}=0.5$.

In Fig.~\ref{fig:Eff_P_mu}(a, b), we find that $\eta_{\mathrm{eng}}/\eta_{\mathrm{C}}$ initially increases with $|\Delta\mu|$.
In Fig.~\ref{fig:Eff_P_mu}(b), as $|\Delta\mu|$ increases further, the chemical potential gradient becomes stronger than the temperature gradient in driving the current and the system stops functioning as an engine.
As a result, efficiency falls quickly to zero. 
The maximum efficiency of the ladder is higher in all regions when $\tilde{E}$ is increased from $\tilde{E}=0.1$ (a) to $0.5$ (b).
This is again similar to the findings in the linear response regime, where we find that $ZT$ (efficiency) increases with $\tilde{E}$. 
In Fig.~\ref{fig:Eff_P_mu}(c, d), we find that the power generated by the ladder follows the same trend as the efficiency.
For both $\tilde{E}=0.1,\ 0.5$, it is possible to improve the efficiency of the ladder without compromising on the power generated.

\section{Conclusions}\label{sec:conclusions}
We have analyzed the thermopower performance of a boundary-driven, non-interacting bosonic ladder in the presence of a gauge field.
Despite being a minimal model, we have shown that its energy band structure can be tuned to deliver a wide range of thermopower performance in linear and nonlinear response regimes.

In the linear response regime, we have studied the maximum efficiency and power of the ladder for various temperatures and chemical potentials.
We have evaluated the importance of band engineering in influencing transport.
We found that both the opening of a bandgap, and the emergence of degenerate ground state play important roles in determining transport.
We found that the emergence of degenerate ground states, which corresponds to the ground state phase transition from the Meissner to vortex phase, influences transport properties at both low and high temperatures.
Away from the linear response regime, we have studied the efficiency and power generated in the ladder while keeping the average temperature or chemical potential constant.
In particular, we have shown how to tune the bath biases to achieve maximum efficiency or power for different regions in the system parameter space.
Depending on if one wishes to maximize efficiency or power, our analysis provides a general guideline on choosing the appropriate system and bath parameters. 
For a wide range of temperatures and chemical potentials, the band structure that features gapped bands with degenerate ground states is the most efficient in converting heat current to power, while at the same time delivering sizable power.

The most convenient candidate to study the setup we proposed is through ultracold bosons in optical lattices. In fact in the past years there have been significant advances in the realization of synthetic gauge fields \cite{DalibardOhberg2011, GoldmanSpielman2014}. Furthermore, the realization of the two-legged ladder with gauge field has already been be achieved in the experimental framework described in~\cite{AtalaBloch2014}.
More specifically, the system can be set up by trapping $^{87}\rm{Rb}$ atoms in a three-dimensional optical lattice potential created by standing waves of different wavelengths in different directions.
The tunneling in the rungs of the ladder are further differentiated by lasers-assisted tunneling.
The bath bias and system-bath interface can be prepared following the description in~\cite{KrinnerEsslinger2017}, where the chemical potential bias of the baths can be tuned by creating unequal populations.
The temperature imbalance can then be introduced by depositing energy into the baths, for instance via heating one of them.

One interesting aspect to explore further is the inclusion of repulsive on-site interaction in the bosonic ladder with gauge field.
This repulsive interaction can be tuned experimentally, for example with a Feshbach resonance~\cite{CourteilleVerhaar1998,RobertsWieman1998,InouyeKetterle1998}, or by varying the local trapping potential~\cite{BlochZwerger2008}.
Such interacting systems are known to exhibit a richer phase diagram, such as vortex-superfluid, Meissner-superfluid, vortex-Mott insulator, and Meissner-Mott insulator phases, depending on the density of the bosons~\cite{PiraudSchollwock2015, GreschnerVekua2015, BuserMeisner2020, JianXue2021}.
It would be thus interesting to explore how the inclusion of on-site interactions can change the thermopower of the ladder.

\section{Acknowledgement}
We acknowledge support from the Ministry of Education of Singapore AcRF MOE Tier-II (Project No. MOE2018-T2-2-142).
We thank G. Benenti for the helpful discussions and the National Supercomputing Center, Singapore (NSCC)~\cite{NSCC} for the computational work which made this paper possible.
\clearpage
\bibliography{main}

\begin{thebibliography}{67}%
\makeatletter
\providecommand \@ifxundefined [1]{%
 \@ifx{#1\undefined}
}%
\providecommand \@ifnum [1]{%
 \ifnum #1\expandafter \@firstoftwo
 \else \expandafter \@secondoftwo
 \fi
}%
\providecommand \@ifx [1]{%
 \ifx #1\expandafter \@firstoftwo
 \else \expandafter \@secondoftwo
 \fi
}%
\providecommand \natexlab [1]{#1}%
\providecommand \enquote  [1]{``#1''}%
\providecommand \bibnamefont  [1]{#1}%
\providecommand \bibfnamefont [1]{#1}%
\providecommand \citenamefont [1]{#1}%
\providecommand \href@noop [0]{\@secondoftwo}%
\providecommand \href [0]{\begingroup \@sanitize@url \@href}%
\providecommand \@href[1]{\@@startlink{#1}\@@href}%
\providecommand \@@href[1]{\endgroup#1\@@endlink}%
\providecommand \@sanitize@url [0]{\catcode `\\12\catcode `\$12\catcode
  `\&12\catcode `\#12\catcode `\^12\catcode `\_12\catcode `\%12\relax}%
\providecommand \@@startlink[1]{}%
\providecommand \@@endlink[0]{}%
\providecommand \url  [0]{\begingroup\@sanitize@url \@url }%
\providecommand \@url [1]{\endgroup\@href {#1}{\urlprefix }}%
\providecommand \urlprefix  [0]{URL }%
\providecommand \Eprint [0]{\href }%
\providecommand \doibase [0]{http://dx.doi.org/}%
\providecommand \selectlanguage [0]{\@gobble}%
\providecommand \bibinfo  [0]{\@secondoftwo}%
\providecommand \bibfield  [0]{\@secondoftwo}%
\providecommand \translation [1]{[#1]}%
\providecommand \BibitemOpen [0]{}%
\providecommand \bibitemStop [0]{}%
\providecommand \bibitemNoStop [0]{.\EOS\space}%
\providecommand \EOS [0]{\spacefactor3000\relax}%
\providecommand \BibitemShut  [1]{\csname bibitem#1\endcsname}%
\let\auto@bib@innerbib\@empty
\bibitem [{\citenamefont {{Mahan}}\ \emph {et~al.}(1997)\citenamefont
  {{Mahan}}, \citenamefont {{Sales}},\ and\ \citenamefont
  {{Sharp}}}]{Mahan1997}%
  \BibitemOpen
  \bibfield  {author} {\bibinfo {author} {\bibfnamefont {G.}~\bibnamefont
  {{Mahan}}}, \bibinfo {author} {\bibfnamefont {B.}~\bibnamefont {{Sales}}}, \
  and\ \bibinfo {author} {\bibfnamefont {J.}~\bibnamefont {{Sharp}}},\ }\href
  {\doibase 10.1063/1.881752} {\bibfield  {journal} {\bibinfo  {journal}
  {Physics Today}\ }\textbf {\bibinfo {volume} {50}},\ \bibinfo {pages} {42}
  (\bibinfo {year} {1997})}\BibitemShut {NoStop}%
\bibitem [{\citenamefont {Dresselhaus}\ \emph {et~al.}(2007)\citenamefont
  {Dresselhaus}, \citenamefont {Chen}, \citenamefont {Tang}, \citenamefont
  {Yang}, \citenamefont {Lee}, \citenamefont {Wang}, \citenamefont {Ren},
  \citenamefont {Fleurial},\ and\ \citenamefont {Gogna}}]{Dresselhaus2007}%
  \BibitemOpen
  \bibfield  {author} {\bibinfo {author} {\bibfnamefont {M.}~\bibnamefont
  {Dresselhaus}}, \bibinfo {author} {\bibfnamefont {G.}~\bibnamefont {Chen}},
  \bibinfo {author} {\bibfnamefont {M.}~\bibnamefont {Tang}}, \bibinfo {author}
  {\bibfnamefont {R.}~\bibnamefont {Yang}}, \bibinfo {author} {\bibfnamefont
  {H.}~\bibnamefont {Lee}}, \bibinfo {author} {\bibfnamefont {D.}~\bibnamefont
  {Wang}}, \bibinfo {author} {\bibfnamefont {Z.}~\bibnamefont {Ren}}, \bibinfo
  {author} {\bibfnamefont {J.-P.}\ \bibnamefont {Fleurial}}, \ and\ \bibinfo
  {author} {\bibfnamefont {P.}~\bibnamefont {Gogna}},\ }\href {\doibase
  https://doi.org/10.1002/adma.200600527} {\bibfield  {journal} {\bibinfo
  {journal} {Advanced Materials}\ }\textbf {\bibinfo {volume} {19}},\ \bibinfo
  {pages} {1043} (\bibinfo {year} {2007})}\BibitemShut {NoStop}%
\bibitem [{\citenamefont {Benenti}\ \emph {et~al.}(2017)\citenamefont
  {Benenti}, \citenamefont {Casati}, \citenamefont {Saito},\ and\ \citenamefont
  {Whitney}}]{BenentiWhitney2017}%
  \BibitemOpen
  \bibfield  {author} {\bibinfo {author} {\bibfnamefont {G.}~\bibnamefont
  {Benenti}}, \bibinfo {author} {\bibfnamefont {G.}~\bibnamefont {Casati}},
  \bibinfo {author} {\bibfnamefont {K.}~\bibnamefont {Saito}}, \ and\ \bibinfo
  {author} {\bibfnamefont {R.~S.}\ \bibnamefont {Whitney}},\ }\href {\doibase
  10.1016/j.physrep.2017.05.008} {\bibfield  {journal} {\bibinfo  {journal}
  {Phys. Rep.}\ }\textbf {\bibinfo {volume} {694}},\ \bibinfo {pages} {1}
  (\bibinfo {year} {2017})}\BibitemShut {NoStop}%
\bibitem [{\citenamefont {Curzon}\ and\ \citenamefont
  {Ahlborn}(1975)}]{Curzon1975EfficiencyOA}%
  \BibitemOpen
  \bibfield  {author} {\bibinfo {author} {\bibfnamefont {F.}~\bibnamefont
  {Curzon}}\ and\ \bibinfo {author} {\bibfnamefont {B.}~\bibnamefont
  {Ahlborn}},\ }\href {\doibase 10.1119/1.10023} {\bibfield  {journal}
  {\bibinfo  {journal} {American Journal of Physics}\ }\textbf {\bibinfo
  {volume} {43}},\ \bibinfo {pages} {22} (\bibinfo {year} {1975})}\BibitemShut
  {NoStop}%
\bibitem [{\citenamefont {Benenti}\ \emph {et~al.}(2011)\citenamefont
  {Benenti}, \citenamefont {Saito},\ and\ \citenamefont
  {Casati}}]{Benenti2011}%
  \BibitemOpen
  \bibfield  {author} {\bibinfo {author} {\bibfnamefont {G.}~\bibnamefont
  {Benenti}}, \bibinfo {author} {\bibfnamefont {K.}~\bibnamefont {Saito}}, \
  and\ \bibinfo {author} {\bibfnamefont {G.}~\bibnamefont {Casati}},\ }\href
  {\doibase 10.1103/PhysRevLett.106.230602} {\bibfield  {journal} {\bibinfo
  {journal} {Phys. Rev. Lett.}\ }\textbf {\bibinfo {volume} {106}},\ \bibinfo
  {pages} {230602} (\bibinfo {year} {2011})}\BibitemShut {NoStop}%
\bibitem [{\citenamefont {Brandner}\ \emph {et~al.}(2013)\citenamefont
  {Brandner}, \citenamefont {Saito},\ and\ \citenamefont
  {Seifert}}]{Brandner2013}%
  \BibitemOpen
  \bibfield  {author} {\bibinfo {author} {\bibfnamefont {K.}~\bibnamefont
  {Brandner}}, \bibinfo {author} {\bibfnamefont {K.}~\bibnamefont {Saito}}, \
  and\ \bibinfo {author} {\bibfnamefont {U.}~\bibnamefont {Seifert}},\ }\href
  {\doibase 10.1103/PhysRevLett.110.070603} {\bibfield  {journal} {\bibinfo
  {journal} {Phys. Rev. Lett.}\ }\textbf {\bibinfo {volume} {110}},\ \bibinfo
  {pages} {070603} (\bibinfo {year} {2013})}\BibitemShut {NoStop}%
\bibitem [{\citenamefont {Whitney}(2014)}]{Whitney_2014}%
  \BibitemOpen
  \bibfield  {author} {\bibinfo {author} {\bibfnamefont {R.~S.}\ \bibnamefont
  {Whitney}},\ }\href {\doibase 10.1103/PhysRevLett.112.130601} {\bibfield
  {journal} {\bibinfo  {journal} {Phys. Rev. Lett.}\ }\textbf {\bibinfo
  {volume} {112}},\ \bibinfo {pages} {130601} (\bibinfo {year}
  {2014})}\BibitemShut {NoStop}%
\bibitem [{\citenamefont {Nolas}\ \emph {et~al.}(2001)\citenamefont {Nolas},
  \citenamefont {Sharp},\ and\ \citenamefont
  {Goldsmid}}]{Nolas2001ThermoelectricsBP}%
  \BibitemOpen
  \bibfield  {author} {\bibinfo {author} {\bibfnamefont {G.~S.}\ \bibnamefont
  {Nolas}}, \bibinfo {author} {\bibfnamefont {J.}~\bibnamefont {Sharp}}, \ and\
  \bibinfo {author} {\bibfnamefont {H.~J.}\ \bibnamefont {Goldsmid}},\ }\href
  {https://www.springer.com/gp/book/9783540412458} {\emph {\bibinfo {title}
  {Thermoelectrics : basic principles and new materials developments}}}\
  (\bibinfo  {publisher} {Springer-Verlag Berlin Heidelberg},\ \bibinfo {year}
  {2001})\BibitemShut {NoStop}%
\bibitem [{\citenamefont {He}\ and\ \citenamefont {Tritt}(2017)}]{Heeaak9997}%
  \BibitemOpen
  \bibfield  {author} {\bibinfo {author} {\bibfnamefont {J.}~\bibnamefont
  {He}}\ and\ \bibinfo {author} {\bibfnamefont {T.~M.}\ \bibnamefont {Tritt}},\
  }\href {https://doi.org/10.1126/science.aak9997} {\bibfield  {journal}
  {\bibinfo  {journal} {Science}\ }\textbf {\bibinfo {volume} {357}} (\bibinfo
  {year} {2017})}\BibitemShut {NoStop}%
\bibitem [{\citenamefont {Snyder}\ and\ \citenamefont
  {Snyder}(2017)}]{Snyder2017}%
  \BibitemOpen
  \bibfield  {author} {\bibinfo {author} {\bibfnamefont {G.~J.}\ \bibnamefont
  {Snyder}}\ and\ \bibinfo {author} {\bibfnamefont {A.~H.}\ \bibnamefont
  {Snyder}},\ }\href {\doibase 10.1039/C7EE02007D} {\bibfield  {journal}
  {\bibinfo  {journal} {Energy Environ. Sci.}\ }\textbf {\bibinfo {volume}
  {10}},\ \bibinfo {pages} {2280} (\bibinfo {year} {2017})}\BibitemShut
  {NoStop}%
\bibitem [{\citenamefont {Goldsmid}(2021)}]{Goldsmid2021}%
  \BibitemOpen
  \bibfield  {author} {\bibinfo {author} {\bibfnamefont {H.~J.}\ \bibnamefont
  {Goldsmid}},\ }\href {https://doi.org/10.1080/14686996.2021.1903816}
  {\bibfield  {journal} {\bibinfo  {journal} {Science and Technology of
  Advanced Materials}\ }\textbf {\bibinfo {volume} {22}},\ \bibinfo {pages}
  {280} (\bibinfo {year} {2021})}\BibitemShut {NoStop}%
\bibitem [{\citenamefont {Goldsmid}(2010)}]{GoldsmidBook}%
  \BibitemOpen
  \bibfield  {author} {\bibinfo {author} {\bibfnamefont {H.~J.}\ \bibnamefont
  {Goldsmid}},\ }\href {https://www.springer.com/gp/book/9783642260926} {\emph
  {\bibinfo {title} {Introduction to thermoelectricity}}}\ (\bibinfo
  {publisher} {Springer-Verlag Berlin Heidelberg},\ \bibinfo {year}
  {2010})\BibitemShut {NoStop}%
\bibitem [{\citenamefont {Brantut}\ \emph {et~al.}(2013)\citenamefont
  {Brantut}, \citenamefont {Grenier}, \citenamefont {Meineke}, \citenamefont
  {Stadler}, \citenamefont {Krinner}, \citenamefont {Kollath}, \citenamefont
  {Esslinger},\ and\ \citenamefont {Georges}}]{Brantut713}%
  \BibitemOpen
  \bibfield  {author} {\bibinfo {author} {\bibfnamefont {J.-P.}\ \bibnamefont
  {Brantut}}, \bibinfo {author} {\bibfnamefont {C.}~\bibnamefont {Grenier}},
  \bibinfo {author} {\bibfnamefont {J.}~\bibnamefont {Meineke}}, \bibinfo
  {author} {\bibfnamefont {D.}~\bibnamefont {Stadler}}, \bibinfo {author}
  {\bibfnamefont {S.}~\bibnamefont {Krinner}}, \bibinfo {author} {\bibfnamefont
  {C.}~\bibnamefont {Kollath}}, \bibinfo {author} {\bibfnamefont
  {T.}~\bibnamefont {Esslinger}}, \ and\ \bibinfo {author} {\bibfnamefont
  {A.}~\bibnamefont {Georges}},\ }\href {\doibase 10.1126/science.1242308}
  {\bibfield  {journal} {\bibinfo  {journal} {Science}\ }\textbf {\bibinfo
  {volume} {342}},\ \bibinfo {pages} {713} (\bibinfo {year}
  {2013})}\BibitemShut {NoStop}%
\bibitem [{\citenamefont {Filippone}\ \emph {et~al.}(2016)\citenamefont
  {Filippone}, \citenamefont {Hekking},\ and\ \citenamefont
  {Minguzzi}}]{Filippone2016}%
  \BibitemOpen
  \bibfield  {author} {\bibinfo {author} {\bibfnamefont {M.}~\bibnamefont
  {Filippone}}, \bibinfo {author} {\bibfnamefont {F.}~\bibnamefont {Hekking}},
  \ and\ \bibinfo {author} {\bibfnamefont {A.}~\bibnamefont {Minguzzi}},\
  }\href {\doibase 10.1103/PhysRevA.93.011602} {\bibfield  {journal} {\bibinfo
  {journal} {Phys. Rev. A}\ }\textbf {\bibinfo {volume} {93}},\ \bibinfo
  {pages} {011602(R)} (\bibinfo {year} {2016})}\BibitemShut {NoStop}%
\bibitem [{\citenamefont {Papoular}\ \emph {et~al.}(2016)\citenamefont
  {Papoular}, \citenamefont {Pitaevskii},\ and\ \citenamefont
  {Stringari}}]{Papoular_2016}%
  \BibitemOpen
  \bibfield  {author} {\bibinfo {author} {\bibfnamefont {D.~J.}\ \bibnamefont
  {Papoular}}, \bibinfo {author} {\bibfnamefont {L.~P.}\ \bibnamefont
  {Pitaevskii}}, \ and\ \bibinfo {author} {\bibfnamefont {S.}~\bibnamefont
  {Stringari}},\ }\href {\doibase 10.1103/PhysRevA.94.023622} {\bibfield
  {journal} {\bibinfo  {journal} {Phys. Rev. A}\ }\textbf {\bibinfo {volume}
  {94}},\ \bibinfo {pages} {023622} (\bibinfo {year} {2016})}\BibitemShut
  {NoStop}%
\bibitem [{\citenamefont {Gallego-Marcos}\ \emph {et~al.}(2014)\citenamefont
  {Gallego-Marcos}, \citenamefont {Platero}, \citenamefont {Nietner},
  \citenamefont {Schaller},\ and\ \citenamefont
  {Brandes}}]{Gallego_Marcos_2014}%
  \BibitemOpen
  \bibfield  {author} {\bibinfo {author} {\bibfnamefont {F.}~\bibnamefont
  {Gallego-Marcos}}, \bibinfo {author} {\bibfnamefont {G.}~\bibnamefont
  {Platero}}, \bibinfo {author} {\bibfnamefont {C.}~\bibnamefont {Nietner}},
  \bibinfo {author} {\bibfnamefont {G.}~\bibnamefont {Schaller}}, \ and\
  \bibinfo {author} {\bibfnamefont {T.}~\bibnamefont {Brandes}},\ }\href
  {\doibase 10.1103/PhysRevA.90.033614} {\bibfield  {journal} {\bibinfo
  {journal} {Phys. Rev. A}\ }\textbf {\bibinfo {volume} {90}},\ \bibinfo
  {pages} {033614} (\bibinfo {year} {2014})}\BibitemShut {NoStop}%
\bibitem [{\citenamefont {Bidasyuk}\ \emph {et~al.}(2018)\citenamefont
  {Bidasyuk}, \citenamefont {Weyrauch}, \citenamefont {Momme},\ and\
  \citenamefont {Prikhodko}}]{Bidasyuk_2018}%
  \BibitemOpen
  \bibfield  {author} {\bibinfo {author} {\bibfnamefont {Y.~M.}\ \bibnamefont
  {Bidasyuk}}, \bibinfo {author} {\bibfnamefont {M.}~\bibnamefont {Weyrauch}},
  \bibinfo {author} {\bibfnamefont {M.}~\bibnamefont {Momme}}, \ and\ \bibinfo
  {author} {\bibfnamefont {O.~O.}\ \bibnamefont {Prikhodko}},\ }\href {\doibase
  10.1088/1361-6455/aae022} {\bibfield  {journal} {\bibinfo  {journal} {Journal
  of Physics B: Atomic, Molecular and Optical Physics}\ }\textbf {\bibinfo
  {volume} {51}},\ \bibinfo {pages} {205301} (\bibinfo {year}
  {2018})}\BibitemShut {NoStop}%
\bibitem [{\citenamefont {de~Oliveira}(2018)}]{Oliveira2018}%
  \BibitemOpen
  \bibfield  {author} {\bibinfo {author} {\bibfnamefont {M.~J.}\ \bibnamefont
  {de~Oliveira}},\ }\href {\doibase 10.1103/PhysRevE.97.012105} {\bibfield
  {journal} {\bibinfo  {journal} {Phys. Rev. E}\ }\textbf {\bibinfo {volume}
  {97}},\ \bibinfo {pages} {012105} (\bibinfo {year} {2018})}\BibitemShut
  {NoStop}%
\bibitem [{\citenamefont {Chien}\ \emph {et~al.}(2015)\citenamefont {Chien},
  \citenamefont {Peotta},\ and\ \citenamefont {Di~Ventra}}]{ChienDiVentra2015}%
  \BibitemOpen
  \bibfield  {author} {\bibinfo {author} {\bibfnamefont {C.~C.}\ \bibnamefont
  {Chien}}, \bibinfo {author} {\bibfnamefont {S.}~\bibnamefont {Peotta}}, \
  and\ \bibinfo {author} {\bibfnamefont {M.}~\bibnamefont {Di~Ventra}},\ }\href
  {\doibase 10.1038/nphys3531} {\bibfield  {journal} {\bibinfo  {journal}
  {Nature Physics}\ }\textbf {\bibinfo {volume} {11}},\ \bibinfo {pages} {998}
  (\bibinfo {year} {2015})}\BibitemShut {NoStop}%
\bibitem [{\citenamefont {Fazio}\ and\ \citenamefont {{van der
  Zant}}(2001)}]{FazioVDZant2001}%
  \BibitemOpen
  \bibfield  {author} {\bibinfo {author} {\bibfnamefont {R.}~\bibnamefont
  {Fazio}}\ and\ \bibinfo {author} {\bibfnamefont {H.}~\bibnamefont {{van der
  Zant}}},\ }\href {\doibase https://doi.org/10.1016/S0370-1573(01)00022-9}
  {\bibfield  {journal} {\bibinfo  {journal} {Physics Reports}\ }\textbf
  {\bibinfo {volume} {355}},\ \bibinfo {pages} {235} (\bibinfo {year}
  {2001})}\BibitemShut {NoStop}%
\bibitem [{\citenamefont {Tanzi}\ \emph {et~al.}(2013)\citenamefont {Tanzi},
  \citenamefont {Lucioni}, \citenamefont {Chaudhuri}, \citenamefont {Gori},
  \citenamefont {Kumar}, \citenamefont {D'Errico}, \citenamefont {Inguscio},\
  and\ \citenamefont {Modugno}}]{TanziModugno2013}%
  \BibitemOpen
  \bibfield  {author} {\bibinfo {author} {\bibfnamefont {L.}~\bibnamefont
  {Tanzi}}, \bibinfo {author} {\bibfnamefont {E.}~\bibnamefont {Lucioni}},
  \bibinfo {author} {\bibfnamefont {S.}~\bibnamefont {Chaudhuri}}, \bibinfo
  {author} {\bibfnamefont {L.}~\bibnamefont {Gori}}, \bibinfo {author}
  {\bibfnamefont {A.}~\bibnamefont {Kumar}}, \bibinfo {author} {\bibfnamefont
  {C.}~\bibnamefont {D'Errico}}, \bibinfo {author} {\bibfnamefont
  {M.}~\bibnamefont {Inguscio}}, \ and\ \bibinfo {author} {\bibfnamefont
  {G.}~\bibnamefont {Modugno}},\ }\href {\doibase
  10.1103/PhysRevLett.111.115301} {\bibfield  {journal} {\bibinfo  {journal}
  {Phys. Rev. Lett.}\ }\textbf {\bibinfo {volume} {111}},\ \bibinfo {pages}
  {115301} (\bibinfo {year} {2013})}\BibitemShut {NoStop}%
\bibitem [{\citenamefont {Simpson}\ \emph {et~al.}(2014)\citenamefont
  {Simpson}, \citenamefont {Gangardt}, \citenamefont {Lerner},\ and\
  \citenamefont {Kr\"uger}}]{SimpsonKruger2014}%
  \BibitemOpen
  \bibfield  {author} {\bibinfo {author} {\bibfnamefont {D.~P.}\ \bibnamefont
  {Simpson}}, \bibinfo {author} {\bibfnamefont {D.~M.}\ \bibnamefont
  {Gangardt}}, \bibinfo {author} {\bibfnamefont {I.~V.}\ \bibnamefont
  {Lerner}}, \ and\ \bibinfo {author} {\bibfnamefont {P.}~\bibnamefont
  {Kr\"uger}},\ }\href {\doibase 10.1103/PhysRevLett.112.100601} {\bibfield
  {journal} {\bibinfo  {journal} {Phys. Rev. Lett.}\ }\textbf {\bibinfo
  {volume} {112}},\ \bibinfo {pages} {100601} (\bibinfo {year}
  {2014})}\BibitemShut {NoStop}%
\bibitem [{\citenamefont {Eckel}\ \emph {et~al.}(2016)\citenamefont {Eckel},
  \citenamefont {Lee}, \citenamefont {Jendrzejewski}, \citenamefont {Lobb},
  \citenamefont {Campbell},\ and\ \citenamefont {Hill}}]{EckelHill2016}%
  \BibitemOpen
  \bibfield  {author} {\bibinfo {author} {\bibfnamefont {S.}~\bibnamefont
  {Eckel}}, \bibinfo {author} {\bibfnamefont {J.~G.}\ \bibnamefont {Lee}},
  \bibinfo {author} {\bibfnamefont {F.}~\bibnamefont {Jendrzejewski}}, \bibinfo
  {author} {\bibfnamefont {C.~J.}\ \bibnamefont {Lobb}}, \bibinfo {author}
  {\bibfnamefont {G.~K.}\ \bibnamefont {Campbell}}, \ and\ \bibinfo {author}
  {\bibfnamefont {W.~T.}\ \bibnamefont {Hill}},\ }\href {\doibase
  10.1103/PhysRevA.93.063619} {\bibfield  {journal} {\bibinfo  {journal} {Phys.
  Rev. A}\ }\textbf {\bibinfo {volume} {93}},\ \bibinfo {pages} {063619}
  (\bibinfo {year} {2016})}\BibitemShut {NoStop}%
\bibitem [{\citenamefont {Krinner}\ \emph {et~al.}(2017)\citenamefont
  {Krinner}, \citenamefont {Esslinger},\ and\ \citenamefont
  {Brantut}}]{KrinnerEsslinger2017}%
  \BibitemOpen
  \bibfield  {author} {\bibinfo {author} {\bibfnamefont {S.}~\bibnamefont
  {Krinner}}, \bibinfo {author} {\bibfnamefont {T.}~\bibnamefont {Esslinger}},
  \ and\ \bibinfo {author} {\bibfnamefont {J.-P.}\ \bibnamefont {Brantut}},\
  }\href {\doibase 10.1088/1361-648X/aa74a1} {\bibfield  {journal} {\bibinfo
  {journal} {Journal of Physics: Condensed Matter}\ }\textbf {\bibinfo {volume}
  {29}},\ \bibinfo {pages} {343003} (\bibinfo {year} {2017})},\ \Eprint
  {http://arxiv.org/abs/1706.01085} {1706.01085} \BibitemShut {NoStop}%
\bibitem [{\citenamefont {Kardar}(1986)}]{Kardar1986}%
  \BibitemOpen
  \bibfield  {author} {\bibinfo {author} {\bibfnamefont {M.}~\bibnamefont
  {Kardar}},\ }\href {\doibase 10.1103/PhysRevB.33.3125} {\bibfield  {journal}
  {\bibinfo  {journal} {Phys. Rev. B}\ }\textbf {\bibinfo {volume} {33}},\
  \bibinfo {pages} {3125} (\bibinfo {year} {1986})}\BibitemShut {NoStop}%
\bibitem [{\citenamefont {Nishiyama}(2000)}]{Nishiyama2000}%
  \BibitemOpen
  \bibfield  {author} {\bibinfo {author} {\bibfnamefont {Y.}~\bibnamefont
  {Nishiyama}},\ }\href {\doibase 10.1007/s100510070144} {\bibfield  {journal}
  {\bibinfo  {journal} {Eur. Phys. J. B}\ }\textbf {\bibinfo {volume} {17}},\
  \bibinfo {pages} {295} (\bibinfo {year} {2000})}\BibitemShut {NoStop}%
\bibitem [{\citenamefont {Atala}\ \emph {et~al.}(2014)\citenamefont {Atala},
  \citenamefont {Aidelsburger}, \citenamefont {Lohse}, \citenamefont
  {Barreiro}, \citenamefont {Paredes},\ and\ \citenamefont
  {Bloch}}]{AtalaBloch2014}%
  \BibitemOpen
  \bibfield  {author} {\bibinfo {author} {\bibfnamefont {M.}~\bibnamefont
  {Atala}}, \bibinfo {author} {\bibfnamefont {M.}~\bibnamefont {Aidelsburger}},
  \bibinfo {author} {\bibfnamefont {M.}~\bibnamefont {Lohse}}, \bibinfo
  {author} {\bibfnamefont {J.~T.}\ \bibnamefont {Barreiro}}, \bibinfo {author}
  {\bibfnamefont {B.}~\bibnamefont {Paredes}}, \ and\ \bibinfo {author}
  {\bibfnamefont {I.}~\bibnamefont {Bloch}},\ }\href {\doibase
  10.1038/nphys2998} {\bibfield  {journal} {\bibinfo  {journal} {Nat. Phys.}\
  }\textbf {\bibinfo {volume} {10}},\ \bibinfo {pages} {588} (\bibinfo {year}
  {2014})}\BibitemShut {NoStop}%
\bibitem [{\citenamefont {Landi}\ \emph {et~al.}(2021)\citenamefont {Landi},
  \citenamefont {Poletti},\ and\ \citenamefont {Schaller}}]{LandiSchaller2021}%
  \BibitemOpen
  \bibfield  {author} {\bibinfo {author} {\bibfnamefont {G.~T.}\ \bibnamefont
  {Landi}}, \bibinfo {author} {\bibfnamefont {D.}~\bibnamefont {Poletti}}, \
  and\ \bibinfo {author} {\bibfnamefont {G.}~\bibnamefont {Schaller}},\
  }\href@noop {} {\enquote {\bibinfo {title} {Non-equilibrium boundary driven
  quantum systems: models, methods and properties},}\ } (\bibinfo {year}
  {2021}),\ \Eprint {http://arxiv.org/abs/2104.14350} {arXiv:2104.14350
  [quant-ph]} \BibitemShut {NoStop}%
\bibitem [{\citenamefont {Guo}\ and\ \citenamefont
  {Poletti}(2016)}]{GuoPoletti2016}%
  \BibitemOpen
  \bibfield  {author} {\bibinfo {author} {\bibfnamefont {C.}~\bibnamefont
  {Guo}}\ and\ \bibinfo {author} {\bibfnamefont {D.}~\bibnamefont {Poletti}},\
  }\href {\doibase 10.1103/PhysRevA.94.033610} {\bibfield  {journal} {\bibinfo
  {journal} {Phys. Rev. A}\ }\textbf {\bibinfo {volume} {94}},\ \bibinfo
  {pages} {033610} (\bibinfo {year} {2016})}\BibitemShut {NoStop}%
\bibitem [{\citenamefont {Rivas}\ and\ \citenamefont
  {{Martin-Delgado}}(2017)}]{RivasMartin-Delgado2017}%
  \BibitemOpen
  \bibfield  {author} {\bibinfo {author} {\bibfnamefont {{\'A}.}~\bibnamefont
  {Rivas}}\ and\ \bibinfo {author} {\bibfnamefont {M.~A.}\ \bibnamefont
  {{Martin-Delgado}}},\ }\href {\doibase 10.1038/s41598-017-06722-x} {\bibfield
   {journal} {\bibinfo  {journal} {Sci. Rep.}\ }\textbf {\bibinfo {volume}
  {7}},\ \bibinfo {pages} {6350} (\bibinfo {year} {2017})}\BibitemShut
  {NoStop}%
\bibitem [{\citenamefont {Guo}\ and\ \citenamefont
  {Poletti}(2017)}]{GuoPoletti2017}%
  \BibitemOpen
  \bibfield  {author} {\bibinfo {author} {\bibfnamefont {C.}~\bibnamefont
  {Guo}}\ and\ \bibinfo {author} {\bibfnamefont {D.}~\bibnamefont {Poletti}},\
  }\href {\doibase 10.1103/PhysRevB.96.165409} {\bibfield  {journal} {\bibinfo
  {journal} {Phys. Rev. B}\ }\textbf {\bibinfo {volume} {96}},\ \bibinfo
  {pages} {165409} (\bibinfo {year} {2017})}\BibitemShut {NoStop}%
\bibitem [{\citenamefont {Xing}\ \emph {et~al.}(2020)\citenamefont {Xing},
  \citenamefont {Xu}, \citenamefont {Balachandran},\ and\ \citenamefont
  {Poletti}}]{XingPoletti2020}%
  \BibitemOpen
  \bibfield  {author} {\bibinfo {author} {\bibfnamefont {B.}~\bibnamefont
  {Xing}}, \bibinfo {author} {\bibfnamefont {X.}~\bibnamefont {Xu}}, \bibinfo
  {author} {\bibfnamefont {V.}~\bibnamefont {Balachandran}}, \ and\ \bibinfo
  {author} {\bibfnamefont {D.}~\bibnamefont {Poletti}},\ }\href {\doibase
  10.1103/PhysRevB.102.245433} {\bibfield  {journal} {\bibinfo  {journal}
  {Phys. Rev. B}\ }\textbf {\bibinfo {volume} {102}},\ \bibinfo {pages}
  {245433} (\bibinfo {year} {2020})}\BibitemShut {NoStop}%
\bibitem [{\citenamefont {Mahan}\ and\ \citenamefont {Sofo}(1996)}]{Mahan7436}%
  \BibitemOpen
  \bibfield  {author} {\bibinfo {author} {\bibfnamefont {G.~D.}\ \bibnamefont
  {Mahan}}\ and\ \bibinfo {author} {\bibfnamefont {J.~O.}\ \bibnamefont
  {Sofo}},\ }\href {\doibase 10.1073/pnas.93.15.7436} {\bibfield  {journal}
  {\bibinfo  {journal} {Proceedings of the National Academy of Sciences}\
  }\textbf {\bibinfo {volume} {93}},\ \bibinfo {pages} {7436} (\bibinfo {year}
  {1996})}\BibitemShut {NoStop}%
\bibitem [{\citenamefont {Pei}\ \emph {et~al.}(2012)\citenamefont {Pei},
  \citenamefont {Wang},\ and\ \citenamefont {Snyder}}]{Pei2012}%
  \BibitemOpen
  \bibfield  {author} {\bibinfo {author} {\bibfnamefont {Y.}~\bibnamefont
  {Pei}}, \bibinfo {author} {\bibfnamefont {H.}~\bibnamefont {Wang}}, \ and\
  \bibinfo {author} {\bibfnamefont {G.~J.}\ \bibnamefont {Snyder}},\ }\href
  {\doibase https://doi.org/10.1002/adma.201202919} {\bibfield  {journal}
  {\bibinfo  {journal} {Advanced Materials}\ }\textbf {\bibinfo {volume}
  {24}},\ \bibinfo {pages} {6125} (\bibinfo {year} {2012})}\BibitemShut
  {NoStop}%
\bibitem [{\citenamefont {Witkoske}\ \emph {et~al.}(2017)\citenamefont
  {Witkoske}, \citenamefont {Wang}, \citenamefont {Lundstrom}, \citenamefont
  {Askarpour},\ and\ \citenamefont {Maassen}}]{Witkoske2017}%
  \BibitemOpen
  \bibfield  {author} {\bibinfo {author} {\bibfnamefont {E.}~\bibnamefont
  {Witkoske}}, \bibinfo {author} {\bibfnamefont {X.}~\bibnamefont {Wang}},
  \bibinfo {author} {\bibfnamefont {M.}~\bibnamefont {Lundstrom}}, \bibinfo
  {author} {\bibfnamefont {V.}~\bibnamefont {Askarpour}}, \ and\ \bibinfo
  {author} {\bibfnamefont {J.}~\bibnamefont {Maassen}},\ }\href {\doibase
  10.1063/1.4994696} {\bibfield  {journal} {\bibinfo  {journal} {Journal of
  Applied Physics}\ }\textbf {\bibinfo {volume} {122}},\ \bibinfo {pages}
  {175102} (\bibinfo {year} {2017})}\BibitemShut {NoStop}%
\bibitem [{\citenamefont {Kumarasinghe}\ and\ \citenamefont
  {Neophytou}(2019)}]{Kumarasinghe2019}%
  \BibitemOpen
  \bibfield  {author} {\bibinfo {author} {\bibfnamefont {C.}~\bibnamefont
  {Kumarasinghe}}\ and\ \bibinfo {author} {\bibfnamefont {N.}~\bibnamefont
  {Neophytou}},\ }\href {\doibase 10.1103/PhysRevB.99.195202} {\bibfield
  {journal} {\bibinfo  {journal} {Phys. Rev. B}\ }\textbf {\bibinfo {volume}
  {99}},\ \bibinfo {pages} {195202} (\bibinfo {year} {2019})}\BibitemShut
  {NoStop}%
\bibitem [{\citenamefont {Rudderham}\ and\ \citenamefont
  {Maassen}(2020)}]{Rudderham2020}%
  \BibitemOpen
  \bibfield  {author} {\bibinfo {author} {\bibfnamefont {C.}~\bibnamefont
  {Rudderham}}\ and\ \bibinfo {author} {\bibfnamefont {J.}~\bibnamefont
  {Maassen}},\ }\href {\doibase 10.1063/1.5138651} {\bibfield  {journal}
  {\bibinfo  {journal} {Journal of Applied Physics}\ }\textbf {\bibinfo
  {volume} {127}},\ \bibinfo {pages} {065105} (\bibinfo {year}
  {2020})}\BibitemShut {NoStop}%
\bibitem [{\citenamefont {Zhou}\ \emph {et~al.}(2011)\citenamefont {Zhou},
  \citenamefont {Yang}, \citenamefont {Chen},\ and\ \citenamefont
  {Dresselhaus}}]{Zhou2011}%
  \BibitemOpen
  \bibfield  {author} {\bibinfo {author} {\bibfnamefont {J.}~\bibnamefont
  {Zhou}}, \bibinfo {author} {\bibfnamefont {R.}~\bibnamefont {Yang}}, \bibinfo
  {author} {\bibfnamefont {G.}~\bibnamefont {Chen}}, \ and\ \bibinfo {author}
  {\bibfnamefont {M.~S.}\ \bibnamefont {Dresselhaus}},\ }\href {\doibase
  10.1103/PhysRevLett.107.226601} {\bibfield  {journal} {\bibinfo  {journal}
  {Phys. Rev. Lett.}\ }\textbf {\bibinfo {volume} {107}},\ \bibinfo {pages}
  {226601} (\bibinfo {year} {2011})}\BibitemShut {NoStop}%
\bibitem [{\citenamefont {Jeong}\ \emph {et~al.}(2012)\citenamefont {Jeong},
  \citenamefont {Kim},\ and\ \citenamefont {Lundstrom}}]{Jeong2012}%
  \BibitemOpen
  \bibfield  {author} {\bibinfo {author} {\bibfnamefont {C.}~\bibnamefont
  {Jeong}}, \bibinfo {author} {\bibfnamefont {R.}~\bibnamefont {Kim}}, \ and\
  \bibinfo {author} {\bibfnamefont {M.~S.}\ \bibnamefont {Lundstrom}},\ }\href
  {\doibase 10.1063/1.4727855} {\bibfield  {journal} {\bibinfo  {journal}
  {Journal of Applied Physics}\ }\textbf {\bibinfo {volume} {111}},\ \bibinfo
  {pages} {113707} (\bibinfo {year} {2012})}\BibitemShut {NoStop}%
\bibitem [{\citenamefont {Caroli}\ \emph {et~al.}(1971)\citenamefont {Caroli},
  \citenamefont {Combescot}, \citenamefont {Nozieres},\ and\ \citenamefont
  {{Saint-James}}}]{CaroliSaint-James1971}%
  \BibitemOpen
  \bibfield  {author} {\bibinfo {author} {\bibfnamefont {C.}~\bibnamefont
  {Caroli}}, \bibinfo {author} {\bibfnamefont {R.}~\bibnamefont {Combescot}},
  \bibinfo {author} {\bibfnamefont {P.}~\bibnamefont {Nozieres}}, \ and\
  \bibinfo {author} {\bibfnamefont {D.}~\bibnamefont {{Saint-James}}},\ }\href
  {\doibase 10.1088/0022-3719/4/8/018} {\bibfield  {journal} {\bibinfo
  {journal} {J. Phys. C: Solid State Phys.}\ }\textbf {\bibinfo {volume} {4}},\
  \bibinfo {pages} {916} (\bibinfo {year} {1971})}\BibitemShut {NoStop}%
\bibitem [{\citenamefont {Meir}\ and\ \citenamefont
  {Wingreen}(1992)}]{MeirWingreen1992}%
  \BibitemOpen
  \bibfield  {author} {\bibinfo {author} {\bibfnamefont {Y.}~\bibnamefont
  {Meir}}\ and\ \bibinfo {author} {\bibfnamefont {N.~S.}\ \bibnamefont
  {Wingreen}},\ }\href {\doibase 10.1103/PhysRevLett.68.2512} {\bibfield
  {journal} {\bibinfo  {journal} {Phys. Rev. Lett.}\ }\textbf {\bibinfo
  {volume} {68}},\ \bibinfo {pages} {2512} (\bibinfo {year}
  {1992})}\BibitemShut {NoStop}%
\bibitem [{\citenamefont {Haug}\ and\ \citenamefont
  {Jauho}(2008)}]{HaugJauho2008}%
  \BibitemOpen
  \bibfield  {author} {\bibinfo {author} {\bibfnamefont {H.}~\bibnamefont
  {Haug}}\ and\ \bibinfo {author} {\bibfnamefont {A.-P.}\ \bibnamefont
  {Jauho}},\ }\href {\doibase 10.1007/978-3-540-73564-9} {\emph {\bibinfo
  {title} {Quantum {{Kinetics}} in {{Transport}} and {{Optics}} of
  {{Semiconductors}}}}}\ (\bibinfo  {publisher} {{Springer-Verlag}},\ \bibinfo
  {address} {{Berlin Heidelberg}},\ \bibinfo {year} {2008})\BibitemShut
  {NoStop}%
\bibitem [{\citenamefont {Prociuk}\ \emph {et~al.}(2010)\citenamefont
  {Prociuk}, \citenamefont {Phillips},\ and\ \citenamefont
  {Dunietz}}]{ProciukDunietz2010}%
  \BibitemOpen
  \bibfield  {author} {\bibinfo {author} {\bibfnamefont {A.}~\bibnamefont
  {Prociuk}}, \bibinfo {author} {\bibfnamefont {H.}~\bibnamefont {Phillips}}, \
  and\ \bibinfo {author} {\bibfnamefont {B.~D.}\ \bibnamefont {Dunietz}},\
  }\href {\doibase 10.1088/1742-6596/220/1/012008} {\bibfield  {journal}
  {\bibinfo  {journal} {J. Phys.: Conf. Ser.}\ }\textbf {\bibinfo {volume}
  {220}},\ \bibinfo {pages} {012008} (\bibinfo {year} {2010})}\BibitemShut
  {NoStop}%
\bibitem [{\citenamefont {Aeberhard}(2011)}]{Aeberhard2011}%
  \BibitemOpen
  \bibfield  {author} {\bibinfo {author} {\bibfnamefont {U.}~\bibnamefont
  {Aeberhard}},\ }\href {\doibase 10.1007/s10825-011-0375-6} {\bibfield
  {journal} {\bibinfo  {journal} {J. Comput. Electron.}\ }\textbf {\bibinfo
  {volume} {10}},\ \bibinfo {pages} {394} (\bibinfo {year} {2011})}\BibitemShut
  {NoStop}%
\bibitem [{\citenamefont {Zimbovskaya}\ and\ \citenamefont
  {Pederson}(2011)}]{ZimbovskayaPederson2011}%
  \BibitemOpen
  \bibfield  {author} {\bibinfo {author} {\bibfnamefont {N.~A.}\ \bibnamefont
  {Zimbovskaya}}\ and\ \bibinfo {author} {\bibfnamefont {M.~R.}\ \bibnamefont
  {Pederson}},\ }\href {\doibase 10.1016/j.physrep.2011.08.002} {\bibfield
  {journal} {\bibinfo  {journal} {Phys. Rep.}\ }\textbf {\bibinfo {volume}
  {509}},\ \bibinfo {pages} {1} (\bibinfo {year} {2011})}\BibitemShut {NoStop}%
\bibitem [{\citenamefont {Nikoli{\'c}}\ \emph {et~al.}(2012)\citenamefont
  {Nikoli{\'c}}, \citenamefont {Saha}, \citenamefont {Markussen},\ and\
  \citenamefont {Thygesen}}]{NikolicThygesen2012}%
  \BibitemOpen
  \bibfield  {author} {\bibinfo {author} {\bibfnamefont {B.~K.}\ \bibnamefont
  {Nikoli{\'c}}}, \bibinfo {author} {\bibfnamefont {K.~K.}\ \bibnamefont
  {Saha}}, \bibinfo {author} {\bibfnamefont {T.}~\bibnamefont {Markussen}}, \
  and\ \bibinfo {author} {\bibfnamefont {K.~S.}\ \bibnamefont {Thygesen}},\
  }\href {\doibase 10.1007/s10825-012-0386-y} {\bibfield  {journal} {\bibinfo
  {journal} {J. Comput. Electron.}\ }\textbf {\bibinfo {volume} {11}},\
  \bibinfo {pages} {78} (\bibinfo {year} {2012})}\BibitemShut {NoStop}%
\bibitem [{\citenamefont {Dhar}\ \emph {et~al.}(2012)\citenamefont {Dhar},
  \citenamefont {Saito},\ and\ \citenamefont {H{\"a}nggi}}]{DharHanggi2012}%
  \BibitemOpen
  \bibfield  {author} {\bibinfo {author} {\bibfnamefont {A.}~\bibnamefont
  {Dhar}}, \bibinfo {author} {\bibfnamefont {K.}~\bibnamefont {Saito}}, \ and\
  \bibinfo {author} {\bibfnamefont {P.}~\bibnamefont {H{\"a}nggi}},\ }\href
  {\doibase 10.1103/PhysRevE.85.011126} {\bibfield  {journal} {\bibinfo
  {journal} {Phys. Rev. E}\ }\textbf {\bibinfo {volume} {85}},\ \bibinfo
  {pages} {011126} (\bibinfo {year} {2012})}\BibitemShut {NoStop}%
\bibitem [{\citenamefont {Wang}\ \emph {et~al.}(2014)\citenamefont {Wang},
  \citenamefont {Agarwalla}, \citenamefont {Li},\ and\ \citenamefont
  {Thingna}}]{WangThingna2014}%
  \BibitemOpen
  \bibfield  {author} {\bibinfo {author} {\bibfnamefont {J.-S.}\ \bibnamefont
  {Wang}}, \bibinfo {author} {\bibfnamefont {B.~K.}\ \bibnamefont {Agarwalla}},
  \bibinfo {author} {\bibfnamefont {H.}~\bibnamefont {Li}}, \ and\ \bibinfo
  {author} {\bibfnamefont {J.}~\bibnamefont {Thingna}},\ }\href {\doibase
  10.1007/s11467-013-0340-x} {\bibfield  {journal} {\bibinfo  {journal} {Front.
  Phys.}\ }\textbf {\bibinfo {volume} {9}},\ \bibinfo {pages} {673} (\bibinfo
  {year} {2014})}\BibitemShut {NoStop}%
\bibitem [{\citenamefont {Ryndyk}(2016)}]{Ryndyk2016}%
  \BibitemOpen
  \bibfield  {author} {\bibinfo {author} {\bibfnamefont {D.}~\bibnamefont
  {Ryndyk}},\ }\href {\doibase 10.1007/978-3-319-24088-6} {\emph {\bibinfo
  {title} {Theory of {{Quantum Transport}} at {{Nanoscale}}}}},\ Vol.\ \bibinfo
  {volume} {184}\ (\bibinfo  {publisher} {{Springer International
  Publishing}},\ \bibinfo {address} {{Cham}},\ \bibinfo {year}
  {2016})\BibitemShut {NoStop}%
\bibitem [{Note1()}]{Note1}%
  \BibitemOpen
  \bibinfo {note} {Simulations at $L=128$ have shown that the results obtained
  are consistent with $L=64$.}\BibitemShut {Stop}%
\bibitem [{\citenamefont {Dittrich}\ \emph {et~al.}(1998)\citenamefont
  {Dittrich}, \citenamefont {H{\"a}nggi}, \citenamefont {Kramer}, \citenamefont
  {Sch{\"o}n}, \citenamefont {Ingold},\ and\ \citenamefont
  {Zwerger}}]{DittrichZwerger1998}%
  \BibitemOpen
  \bibfield  {author} {\bibinfo {author} {\bibfnamefont {T.}~\bibnamefont
  {Dittrich}}, \bibinfo {author} {\bibfnamefont {P.}~\bibnamefont
  {H{\"a}nggi}}, \bibinfo {author} {\bibfnamefont {B.}~\bibnamefont {Kramer}},
  \bibinfo {author} {\bibfnamefont {G.}~\bibnamefont {Sch{\"o}n}}, \bibinfo
  {author} {\bibfnamefont {G.-L.}\ \bibnamefont {Ingold}}, \ and\ \bibinfo
  {author} {\bibfnamefont {W.}~\bibnamefont {Zwerger}},\ }\href@noop {} {\emph
  {\bibinfo {title} {Quantum {{Transport}} and {{Dissipation}}}}}\ (\bibinfo
  {publisher} {{Wiley-VCH, Weinheim}},\ \bibinfo {year} {1998})\BibitemShut
  {NoStop}%
\bibitem [{\citenamefont {Landauer}(1957)}]{Landauer1957}%
  \BibitemOpen
  \bibfield  {author} {\bibinfo {author} {\bibfnamefont {R.}~\bibnamefont
  {Landauer}},\ }\href {\doibase 10.1147/rd.13.0223} {\bibfield  {journal}
  {\bibinfo  {journal} {IBM J. Res. Dev.}\ }\textbf {\bibinfo {volume} {1}},\
  \bibinfo {pages} {223} (\bibinfo {year} {1957})}\BibitemShut {NoStop}%
\bibitem [{\citenamefont {Landauer}(1970)}]{Landauer1970}%
  \BibitemOpen
  \bibfield  {author} {\bibinfo {author} {\bibfnamefont {R.}~\bibnamefont
  {Landauer}},\ }\href {\doibase 10.1080/14786437008238472} {\bibfield
  {journal} {\bibinfo  {journal} {Philos. Mag.}\ }\textbf {\bibinfo {volume}
  {21}},\ \bibinfo {pages} {863} (\bibinfo {year} {1970})}\BibitemShut
  {NoStop}%
\bibitem [{\citenamefont {Datta}(1995)}]{Datta1995}%
  \BibitemOpen
  \bibfield  {author} {\bibinfo {author} {\bibfnamefont {S.}~\bibnamefont
  {Datta}},\ }\href {\doibase 10.1017/CBO9780511805776} {\emph {\bibinfo
  {title} {Electronic {{Transport}} in {{Mesoscopic Systems}}}}}\ (\bibinfo
  {publisher} {{Cambridge University Press}},\ \bibinfo {address}
  {{Cambridge}},\ \bibinfo {year} {1995})\BibitemShut {NoStop}%
\bibitem [{Note2()}]{Note2}%
  \BibitemOpen
  \bibinfo {note} {We benchmarked our transmission function against the ones
  obtained from Kwant~\cite {groth_kwant_2014}, a widely used and cited python
  package for quantum transport, and found perfect agreement.}\BibitemShut
  {Stop}%
\bibitem [{\citenamefont {Dalibard}\ \emph {et~al.}(2011)\citenamefont
  {Dalibard}, \citenamefont {Gerbier}, \citenamefont {Juzeli{\=u}nas},\ and\
  \citenamefont {{\"O}hberg}}]{DalibardOhberg2011}%
  \BibitemOpen
  \bibfield  {author} {\bibinfo {author} {\bibfnamefont {J.}~\bibnamefont
  {Dalibard}}, \bibinfo {author} {\bibfnamefont {F.}~\bibnamefont {Gerbier}},
  \bibinfo {author} {\bibfnamefont {G.}~\bibnamefont {Juzeli{\=u}nas}}, \ and\
  \bibinfo {author} {\bibfnamefont {P.}~\bibnamefont {{\"O}hberg}},\ }\href
  {\doibase 10.1103/RevModPhys.83.1523} {\bibfield  {journal} {\bibinfo
  {journal} {Rev. Mod. Phys.}\ }\textbf {\bibinfo {volume} {83}},\ \bibinfo
  {pages} {1523} (\bibinfo {year} {2011})}\BibitemShut {NoStop}%
\bibitem [{\citenamefont {Goldman}\ \emph {et~al.}(2014)\citenamefont
  {Goldman}, \citenamefont {Juzeli{\=u}nas}, \citenamefont {{\"O}hberg},\ and\
  \citenamefont {Spielman}}]{GoldmanSpielman2014}%
  \BibitemOpen
  \bibfield  {author} {\bibinfo {author} {\bibfnamefont {N.}~\bibnamefont
  {Goldman}}, \bibinfo {author} {\bibfnamefont {G.}~\bibnamefont
  {Juzeli{\=u}nas}}, \bibinfo {author} {\bibfnamefont {P.}~\bibnamefont
  {{\"O}hberg}}, \ and\ \bibinfo {author} {\bibfnamefont {I.~B.}\ \bibnamefont
  {Spielman}},\ }\href {\doibase 10.1088/0034-4885/77/12/126401} {\bibfield
  {journal} {\bibinfo  {journal} {Rep. Prog. Phys.}\ }\textbf {\bibinfo
  {volume} {77}},\ \bibinfo {pages} {126401} (\bibinfo {year}
  {2014})}\BibitemShut {NoStop}%
\bibitem [{\citenamefont {Courteille}\ \emph {et~al.}(1998)\citenamefont
  {Courteille}, \citenamefont {Freeland}, \citenamefont {Heinzen},
  \citenamefont {van Abeelen},\ and\ \citenamefont
  {Verhaar}}]{CourteilleVerhaar1998}%
  \BibitemOpen
  \bibfield  {author} {\bibinfo {author} {\bibfnamefont {P.}~\bibnamefont
  {Courteille}}, \bibinfo {author} {\bibfnamefont {R.~S.}\ \bibnamefont
  {Freeland}}, \bibinfo {author} {\bibfnamefont {D.~J.}\ \bibnamefont
  {Heinzen}}, \bibinfo {author} {\bibfnamefont {F.~A.}\ \bibnamefont {van
  Abeelen}}, \ and\ \bibinfo {author} {\bibfnamefont {B.~J.}\ \bibnamefont
  {Verhaar}},\ }\href {\doibase 10.1103/PhysRevLett.81.69} {\bibfield
  {journal} {\bibinfo  {journal} {Phys. Rev. Lett.}\ }\textbf {\bibinfo
  {volume} {81}},\ \bibinfo {pages} {69} (\bibinfo {year} {1998})}\BibitemShut
  {NoStop}%
\bibitem [{\citenamefont {Roberts}\ \emph {et~al.}(1998)\citenamefont
  {Roberts}, \citenamefont {Claussen}, \citenamefont {Burke}, \citenamefont
  {Greene}, \citenamefont {Cornell},\ and\ \citenamefont
  {Wieman}}]{RobertsWieman1998}%
  \BibitemOpen
  \bibfield  {author} {\bibinfo {author} {\bibfnamefont {J.~L.}\ \bibnamefont
  {Roberts}}, \bibinfo {author} {\bibfnamefont {N.~R.}\ \bibnamefont
  {Claussen}}, \bibinfo {author} {\bibfnamefont {J.~P.}\ \bibnamefont {Burke}},
  \bibinfo {author} {\bibfnamefont {C.~H.}\ \bibnamefont {Greene}}, \bibinfo
  {author} {\bibfnamefont {E.~A.}\ \bibnamefont {Cornell}}, \ and\ \bibinfo
  {author} {\bibfnamefont {C.~E.}\ \bibnamefont {Wieman}},\ }\href {\doibase
  10.1103/PhysRevLett.81.5109} {\bibfield  {journal} {\bibinfo  {journal}
  {Phys. Rev. Lett.}\ }\textbf {\bibinfo {volume} {81}},\ \bibinfo {pages}
  {5109} (\bibinfo {year} {1998})}\BibitemShut {NoStop}%
\bibitem [{\citenamefont {Inouye}\ \emph {et~al.}(1998)\citenamefont {Inouye},
  \citenamefont {Andrews}, \citenamefont {Stenger}, \citenamefont {Miesner},
  \citenamefont {Stamper-Kurn},\ and\ \citenamefont
  {Ketterle}}]{InouyeKetterle1998}%
  \BibitemOpen
  \bibfield  {author} {\bibinfo {author} {\bibfnamefont {S.}~\bibnamefont
  {Inouye}}, \bibinfo {author} {\bibfnamefont {M.~R.}\ \bibnamefont {Andrews}},
  \bibinfo {author} {\bibfnamefont {J.}~\bibnamefont {Stenger}}, \bibinfo
  {author} {\bibfnamefont {H.-J.}\ \bibnamefont {Miesner}}, \bibinfo {author}
  {\bibfnamefont {D.~M.}\ \bibnamefont {Stamper-Kurn}}, \ and\ \bibinfo
  {author} {\bibfnamefont {W.}~\bibnamefont {Ketterle}},\ }\href {\doibase
  10.1038/32354} {\bibfield  {journal} {\bibinfo  {journal} {Nature}\ }\textbf
  {\bibinfo {volume} {392}},\ \bibinfo {pages} {151} (\bibinfo {year}
  {1998})}\BibitemShut {NoStop}%
\bibitem [{\citenamefont {Bloch}\ \emph {et~al.}(2008)\citenamefont {Bloch},
  \citenamefont {Dalibard},\ and\ \citenamefont {Zwerger}}]{BlochZwerger2008}%
  \BibitemOpen
  \bibfield  {author} {\bibinfo {author} {\bibfnamefont {I.}~\bibnamefont
  {Bloch}}, \bibinfo {author} {\bibfnamefont {J.}~\bibnamefont {Dalibard}}, \
  and\ \bibinfo {author} {\bibfnamefont {W.}~\bibnamefont {Zwerger}},\ }\href
  {\doibase 10.1103/RevModPhys.80.885} {\bibfield  {journal} {\bibinfo
  {journal} {Rev. Mod. Phys.}\ }\textbf {\bibinfo {volume} {80}},\ \bibinfo
  {pages} {885} (\bibinfo {year} {2008})}\BibitemShut {NoStop}%
\bibitem [{\citenamefont {Piraud}\ \emph {et~al.}(2015)\citenamefont {Piraud},
  \citenamefont {{Heidrich-Meisner}}, \citenamefont {McCulloch}, \citenamefont
  {Greschner}, \citenamefont {Vekua},\ and\ \citenamefont
  {Schollw{\"o}ck}}]{PiraudSchollwock2015}%
  \BibitemOpen
  \bibfield  {author} {\bibinfo {author} {\bibfnamefont {M.}~\bibnamefont
  {Piraud}}, \bibinfo {author} {\bibfnamefont {F.}~\bibnamefont
  {{Heidrich-Meisner}}}, \bibinfo {author} {\bibfnamefont {I.~P.}\ \bibnamefont
  {McCulloch}}, \bibinfo {author} {\bibfnamefont {S.}~\bibnamefont
  {Greschner}}, \bibinfo {author} {\bibfnamefont {T.}~\bibnamefont {Vekua}}, \
  and\ \bibinfo {author} {\bibfnamefont {U.}~\bibnamefont {Schollw{\"o}ck}},\
  }\href {\doibase 10.1103/PhysRevB.91.140406} {\bibfield  {journal} {\bibinfo
  {journal} {Phys. Rev. B}\ }\textbf {\bibinfo {volume} {91}},\ \bibinfo
  {pages} {140406(R)} (\bibinfo {year} {2015})}\BibitemShut {NoStop}%
\bibitem [{\citenamefont {Greschner}\ \emph {et~al.}(2015)\citenamefont
  {Greschner}, \citenamefont {Piraud}, \citenamefont {{Heidrich-Meisner}},
  \citenamefont {McCulloch}, \citenamefont {Schollw{\"o}ck},\ and\
  \citenamefont {Vekua}}]{GreschnerVekua2015}%
  \BibitemOpen
  \bibfield  {author} {\bibinfo {author} {\bibfnamefont {S.}~\bibnamefont
  {Greschner}}, \bibinfo {author} {\bibfnamefont {M.}~\bibnamefont {Piraud}},
  \bibinfo {author} {\bibfnamefont {F.}~\bibnamefont {{Heidrich-Meisner}}},
  \bibinfo {author} {\bibfnamefont {I.~P.}\ \bibnamefont {McCulloch}}, \bibinfo
  {author} {\bibfnamefont {U.}~\bibnamefont {Schollw{\"o}ck}}, \ and\ \bibinfo
  {author} {\bibfnamefont {T.}~\bibnamefont {Vekua}},\ }\href {\doibase
  10.1103/PhysRevLett.115.190402} {\bibfield  {journal} {\bibinfo  {journal}
  {Phys. Rev. Lett.}\ }\textbf {\bibinfo {volume} {115}},\ \bibinfo {pages}
  {190402} (\bibinfo {year} {2015})}\BibitemShut {NoStop}%
\bibitem [{\citenamefont {Buser}\ \emph {et~al.}(2020)\citenamefont {Buser},
  \citenamefont {Hubig}, \citenamefont {Schollw\"ock}, \citenamefont
  {Tarruell},\ and\ \citenamefont {Heidrich-Meisner}}]{BuserMeisner2020}%
  \BibitemOpen
  \bibfield  {author} {\bibinfo {author} {\bibfnamefont {M.}~\bibnamefont
  {Buser}}, \bibinfo {author} {\bibfnamefont {C.}~\bibnamefont {Hubig}},
  \bibinfo {author} {\bibfnamefont {U.}~\bibnamefont {Schollw\"ock}}, \bibinfo
  {author} {\bibfnamefont {L.}~\bibnamefont {Tarruell}}, \ and\ \bibinfo
  {author} {\bibfnamefont {F.}~\bibnamefont {Heidrich-Meisner}},\ }\href
  {\doibase 10.1103/PhysRevA.102.053314} {\bibfield  {journal} {\bibinfo
  {journal} {Phys. Rev. A}\ }\textbf {\bibinfo {volume} {102}},\ \bibinfo
  {pages} {053314} (\bibinfo {year} {2020})}\BibitemShut {NoStop}%
\bibitem [{\citenamefont {Jian}\ \emph {et~al.}(2021)\citenamefont {Jian},
  \citenamefont {Qiao}, \citenamefont {Liang}, \citenamefont {Yu},
  \citenamefont {Zhang},\ and\ \citenamefont {Xue}}]{JianXue2021}%
  \BibitemOpen
  \bibfield  {author} {\bibinfo {author} {\bibfnamefont {Y.}~\bibnamefont
  {Jian}}, \bibinfo {author} {\bibfnamefont {X.}~\bibnamefont {Qiao}}, \bibinfo
  {author} {\bibfnamefont {J.-C.}\ \bibnamefont {Liang}}, \bibinfo {author}
  {\bibfnamefont {Z.-F.}\ \bibnamefont {Yu}}, \bibinfo {author} {\bibfnamefont
  {A.-X.}\ \bibnamefont {Zhang}}, \ and\ \bibinfo {author} {\bibfnamefont
  {J.-K.}\ \bibnamefont {Xue}},\ }\href {\doibase 10.1103/PhysRevE.104.024212}
  {\bibfield  {journal} {\bibinfo  {journal} {Phys. Rev. E}\ }\textbf {\bibinfo
  {volume} {104}},\ \bibinfo {pages} {024212} (\bibinfo {year}
  {2021})}\BibitemShut {NoStop}%
\bibitem [{NSC()}]{NSCC}%
  \BibitemOpen
  \href@noop {} {}\bibinfo {howpublished} {https://www.nscc.sg/}\BibitemShut
  {NoStop}%
\bibitem [{\citenamefont {Groth}\ \emph {et~al.}(2014)\citenamefont {Groth},
  \citenamefont {Wimmer}, \citenamefont {Akhmerov},\ and\ \citenamefont
  {Waintal}}]{groth_kwant_2014}%
  \BibitemOpen
  \bibfield  {author} {\bibinfo {author} {\bibfnamefont {C.~W.}\ \bibnamefont
  {Groth}}, \bibinfo {author} {\bibfnamefont {M.}~\bibnamefont {Wimmer}},
  \bibinfo {author} {\bibfnamefont {A.~R.}\ \bibnamefont {Akhmerov}}, \ and\
  \bibinfo {author} {\bibfnamefont {X.}~\bibnamefont {Waintal}},\ }\href
  {\doibase 10.1088/1367-2630/16/6/063065} {\bibfield  {journal} {\bibinfo
  {journal} {New Journal of Physics}\ }\textbf {\bibinfo {volume} {16}},\
  \bibinfo {pages} {063065} (\bibinfo {year} {2014})}\BibitemShut {NoStop}%
\end{thebibliography}%

\end{document}